\documentclass[preprint,secnumaRabic,amssymb, nobibnotes, aps, pra, superscriptaddress,longbibliography]{revtex4-2}
\usepackage{graphicx}
\usepackage{braket}
\usepackage{array}
\usepackage{amsmath}
\usepackage{here}
\usepackage{sidecap}
\usepackage{siunitx}
\usepackage{textcomp}
\usepackage{pgfplots}

\begin{document}

\title{Evaluating states in trapped ions with local correlation between internal and motional degrees of freedom}

\author{Silpa Muralidharan}%
\email{u928298g@ecs.osaka-u.ac.jp}
\affiliation{Graduate School of Engineering Science, Osaka University, 1-3 Machikaneyama, Toyonaka, Osaka, Japan}
\author{Ryutaro Ohira}%
\affiliation{Graduate School of Engineering Science, Osaka University, 1-3 Machikaneyama, Toyonaka, Osaka, Japan}
\affiliation{Center for Quantum Information and Quantum Biology, Osaka University, 1-2 Machikaneyama, Toyonaka, Osaka, Japan}
\author{Shota Kume}%
\affiliation{Graduate School of Engineering Science, Osaka University, 1-3 Machikaneyama, Toyonaka, Osaka, Japan}
\author{Kenji Toyoda}%
\email{toyoda@qiqb.otri.osaka-u.ac.jp}
\affiliation{Center for Quantum Information and Quantum Biology,  Osaka University, 1-2 Machikaneyama, Toyonaka, Osaka, Japan}

\date{\today}

\begin{abstract}

We propose and demonstrate a scalable scheme for 
the simultaneous determination of internal and motional states 
in trapped ions with single-site resolution.
The scheme is applied to the study of
polaritonic excitations in the
Jaynes--Cummings--Hubbard model with trapped ions,
in which the internal and motional states of the ions are strongly correlated. 
We observe quantum phase transitions of polaritonic excitations in two ions by directly evaluating their variances per ion site. Our work establishes an essential technological method for large-scale quantum simulations of polaritonic systems.

\end{abstract}

\maketitle
\section{Introduction}
\label{sec:intro}
Quantum simulations allow us to study the properties of many-body quantum 
systems 
that are hard to investigate with classical computers
\cite{feynman1982}. 
A promising platform for realizing quantum simulations is 
a system of trapped ions \cite{Blatt2012,Monroe2019}. Trapped ions have advantages over other systems from such perspectives as ease of preparation and control, and we can address individual particles with little perturbation to neighboring particles.

The Jaynes--Cummings  (JC) model describes the  atom--field interaction in
a combined system  of a two-level atom and  a quantized electric-field
mode \cite{Jaynes1963}. 
An interconnected array of two-level atoms interacting with quantized wave modes is known as 
the Jaynes--Cummings--Hubbard (JCH) model,
and related systems of coupled cavity arrays have been extensively investigated 
\cite{Greentree2006,hartmann2006,Angelakis2007,Hartmann2008,PhysRevA.85.013810,Gessner2015,Hartmann2016}.

The JCH model can be realized with arrays of cavity QED systems \cite{Chang2018,Kato2019}
and of circuit QED systems \cite{Hoffman2011,Houck2012,Schmidt2013,Raftery2014,Fitzpatrick2017}, as well as
systems of trapped ions \cite{Ivanov2009,Toyo2013,Gessner2015,PhysRevLett.120.073001,PhysRevA.103.012612,ohira2021polariton}. 
These systems can be flexibly controlled with sets of system parameters. Therefore, they can be considered to be attractive systems
for studying quantum many-body phenomena.

In the JCH model, quasi-particles called polaritons play an essential role. Each polariton is represented as the superposition of an internal excitation and a photon or a motional excitation (phonon). Polaritons are well-defined particles in the JCH model and their total number is conserved. The implications for  understanding the JCH model is apparent in the quantum phase transition of a JCH system between the Mott-insulator (MI) and superfluid (SF) phases, which are characterized by a drastic change in the polariton-number variance per site \cite{Angelakis2007}.

In the MI phase, each polariton is fixed to a respective site, and fluctuations of the polariton number per site vanish. In the SF phase, each polariton is distributed over multiple sites, and the ensemble of polaritons forms a coherent wave-like state that spreads over the entire system. In the latter case, the variance of the polariton number per site becomes finite. Therefore, by evaluating the polariton-number variance per site while the system parameters are varied, the quantum phase transition from the MI phase to the SF phase can be detected. In the case of the JCH model with trapped ions \cite{Ivanov2009}, each polariton is a correlated combination of an internal excitation and a vibrational quantum (local phonon \cite{Porras2004, Brown2011, Harlander2011, Haze2012,wilson2014tunable,toyoda2015hong,PhysRevLett.120.073001,PhysRevLett.124.200501,PhysRevA.100.060301,PhysRevA.103.012612}). Therefore, by detecting the internal and motional states, the states of polaritons can be precisely evaluated. It is not easy to directly measure the motional states and they should instead be first mapped to the internal states to be detected by, e.g.,  state-dependent fluorescence detection.
However, this process destroys the internal states.
The internal and motional states cannot be simultaneously determined 
even if we measure the former and the latter consecutively in this order, 
since the measurement of the internal states with fluorescence detection may destroy the motional states due to photon recoil during several fluorescence cycles.

One approach to address this is a conditional measurement method    \cite{Ivanov2009, Wang2010} using levels in such ions as $^{40}$Ca$^+$, which has a metastable state. In this method, a random guess of the internal states (the ground state $\ket{g}$ or the metastable state $\ket{e}$) of a particular ion is first made. If the guess is made for the $\ket{g}$ or $\ket{e}$ state, a carrier $\pi$ pulse or no pulse, respectively, is applied to the $\ket{g}$--$\ket{e}$ transition. Then, laser light resonant to the transition between $\ket{g}$ and a short-lived excited state is exposed to the ions.  If the ion scatters fluorescence at this point, the initial guess was wrong, and the measurement result is discarded. Otherwise, the initial guess was right, and the internal state is subsequently transferred back to the $\ket{g}$ state. An analysis of the motional states using a blue-sideband (BSB) Rabi oscillation is then performed. Fourier analyses on a series of these results give information on the motional states and the internal states that correspond to the initial guesses. This combined information retains the local correlation between the internal and motional states in the states of interest. This method can be  applied to the case of a single ion.

In the case of an ion chain, the concept of local vibrational modes (local phonons) becomes approximate. Each mode can be coupled with those in different ion sites through Coulomb couplings. When a particular ion site is heated due to photon recoils during fluorescence cycles, the heat is transferred to different ion sites, and eventually the motional states at those sites may be destroyed (mutual heating among ion sites).

In this article, we demonstrate the measurement of local motional states conditioned on the internal states by shelving part of the ions to a long-lived internal state. 
In this method, fluorescence cycles are not used to determine the internal states, and the probability amplitudes that are of no interest
are simply hidden in the auxiliary long-lived internal state.
Therefore, this method does not suffer from mutual heating among ion sites as explained above, and is applicable to the measurement of local motional states in a multi-ion crystal.
In this work, this method is applied to determine the polariton number and its variance in the JCH model while the system parameters are changed.  The method is scalable with respect to the number of ion sites, i.e., the same time sequence can be used in principle for an arbitrary number of ions in the chain.

\section{JCH model}
\label{sec:jchmodel}

The JC model \cite{Jaynes1963} was initially proposed by E. Jaynes and F. Cummings in 1963. It is a quantum optics model that describes the interaction of a two-level atom with a single quantized mode of an optical cavity electromagnetic field. The model is widely used in such fields as cavity QED and circuit QED.
The Hamiltonian for the JC model is given as follows:
\begin{equation}
    H_\mathrm{JC} = \hbar \omega \hat{a}^\dagger \hat{a} +\frac{1}{2}\hbar \omega_0\hat{\sigma}_z + \hbar g (\hat{\sigma}_+\hat{a} + \hat{\sigma}_- \hat{a}^\dagger),
\end{equation}
where $\omega$ is the oscillation frequency of the electromagnetic field, 
$\hat{a}^\dagger$, $\hat{a}$  are the creation and annihilation operators for the quantized mode of the electromagnetic field, 
$\omega_0$ is the resonance frequency of the two-level atom, 
$\hat{\sigma}_z$ is the Pauli operator in the $z$ direction,
$g$ is the JC coupling coefficient and
$\hat{\sigma}_+$, $\hat{\sigma}_-$ are the atomic raising and lowering operators. 
This model introduces superposition states corresponding to dressed atoms or polaritons and a semi-infinite series of eigenenergies called the JC ladder \cite{Jaynes1963}. It leads to such phenomena as a vacuum-field Rabi oscillation or collapses and revivals of atomic oscillations. 
The analogy between a cavity QED system and a trapped-ion system leads to implementing the JC model in trapped-ion experiments for quantum information processing.

When trapped ions in an ion chain with relatively tight confinement along the radial directions \cite{Porras2004}  are used and optical fields resonant to the red-sideband (RSB) transitions are applied, the two-level systems in the ions interacting with radial local phonons can be considered as an interconnected array of JC systems, which can be described with the JCH model.
The Hamiltonian for the JCH model in a system of two trapped ions is given as follows \cite{Ivanov2009}:

\begin{eqnarray}
    H_\mathrm{JCH} &=&
\hbar \frac{\kappa}{2} ( \hat{a}^\dagger_1 \hat{a}_2
+  \hat{a}_1\hat{a}^\dagger_2) 
 \nonumber
+   \hbar \Delta \sum_{j=1,2} \ket{e_j}\bra{e_j} \\
&&+ \hbar g\sum_{j=1,2} 
(\hat{a}^\dagger_j \hat{\sigma}^{-}_j +\hat{a}_j \hat{\sigma}^{+}_j), 
\end{eqnarray}
where 
$\kappa$ is the hopping rate of the radial local phonons,
$\hat{a}^\dagger_j$ and $\hat{a}_j$ are the creation and annihilation operators for phonons at the $j$th ion, 
$\Delta$ is the amount of detuning of the optical field from the RSB transition, 
$\ket{e_j}$ is the internal excited state in the $j$th ion,
and
${\sigma}^{+}_j \equiv \ket{e_j}\bra{g_j} $, ${\sigma}^{-}_j \equiv \ket{g_j}\bra{e_j} $ 
are the raising and lowering operators for the internal states
in the $j$th ion,
where $\ket{g_j}$ is the internal ground state in the $j$th ion.
In this JCH  system, quantum phase transitions between the MI and SF phases are expected to occur \cite{Greentree2006,Angelakis2007}. A quantum phase transition is similar to a classical phase transition, while the cause of the transition is directly related to quantum fluctuations instead of such parameters as pressure or temperature.

\section{MI-to-SF quantum phase transitions in the JCH model}
\label{sec:misf}

In the JCH model, a quantum phase transition from the MI phase to the SF phase in a trapped-ion chain is realized by the adiabatic transfer process \cite{hartmann2006, Greentree2006}.
Here, we consider that the ions are prepared in 
$\ket{\psi_\mathrm{MI}}=\ket{e_1}\ket{e_2}\ket{0}_1\ket{0}_2$. 
This state corresponds to a ground state, the MI state, in which the excitations are localized.
The parameters $\Delta$ and $g$ can be changed by varying the frequency and amplitude of the laser, respectively.
We linearly sweep $\Delta$ from a negative value to a positive one and change the amplitude (and hence $g$) following a Gaussian shape. 
In this case, the Hamiltonian changes time-dependently 
and an adiabatic transfer is induced \cite{Nikolay2001}. At the final stage of this transfer, a ground state called the SF phase is created which has the form
$\ket{\psi_\mathrm{SF}} = \ket{g_1}\ket{g_2} \otimes [(1/\sqrt{2})\ket{1}_1\ket{1}_2 
-(1/2       )\ket{2}_1\ket{0}_2
-(1/2       )\ket{0}_1\ket{2}_2 ]$
\cite{Irish2008}.
In the intermediate region of the adiabatic transfer, polaritonic MI and SF phases exist.

In an analogy to second-order phase transitions in statistical physics, 
quantum phase transitions in the current system
can be understood by the emergence and disappearance of certain ordered phases.
The MI states are considered to have an order in the excitation
(polariton) number
per site, in close analogy to the Bose--Hubbard systems \cite{Jaksch1998,greiner2002}.
The SF states in this system are believed to have coherence among multiple sites.
In the case of two ions, the phonon and polaritonic SF states 
can be represented as superpositions of product states over multiple 
sites \cite{Irish2008}.
Although it is not straightforward to define an order parameter
that embodies the coherence among multiple sites,
the order in the polariton number per site can be readily defined.

Angelakis {\it et al.} proposed using the polariton-number variance per site
to determine the different phases of the system \cite{Angelakis2007};
it tends to be zero for the MI phase, which has perfect number order,
while it takes finite values for the SF phase.
Accordingly, we use the variance for the total excitation number (polariton number) per site to determine quantum phase transitions in the JCH model.
Other excitation-number variances (atomic and phonon) can be used for 
purposes such as classifying the ground states of the system. 

The variance for the total excitation number per site (total variance) is given by $\Delta \hat{N}_j^2\equiv \langle \hat{N}_{j}^2\rangle - \langle \hat{N}_{j}\rangle ^2 $, where $\hat{N}_j = \ket{e_j}\bra{e_j}+\hat{a}_j^{\dagger}\hat{a}_j$. 
The variance for the atomic excitation number per site (atomic variance) is given by $\Delta \hat{N}_{a,j}^2\equiv \langle \hat{N}_{a,j}^2\rangle - \langle \hat{N}_{a,j}\rangle ^2$, where $\hat{N}_{a,j} = \ket{e_j} \bra{e_j}$. 
This helps to identify polaritons. 
Additionally, 
the variance for the phonon number per site (phonon variance) is defined as $\Delta \hat{N}_{p,j}^2\equiv \langle \hat{N}_{p,j}^2\rangle - \langle \hat{N}_{p,j}\rangle ^2$, where $\hat{N}_{p,j} = \hat{a}_j^{\dagger}\hat{a}_j$. 
This follows an increasing trend similar to that for the total variance
in MI-to-SF quantum phase transitions, while its detailed behavior in
the intermediate region is different from that for the total variance.
 The values for the total and atomic variances and the corresponding classes of the ground states are summarized in Table~\ref{tab:1}. 

\begin{table}[h]
\begin{tabular}{ m{2.5cm} m{2.5cm} m{2.5cm} }
\hline
\textbf{} &\textbf{$\Delta N_{a,k} =0$} &\textbf{$\Delta N_{a,k} \neq0$} \\
\hline\hline $\Delta N_{t,k} =0$ & Atomic MI & Polaritonic MI \\
$\Delta N_{t,k} \neq 0$ & Phonon SF & Polaritonic SF \\
\hline
\end{tabular}
\caption{Variances and classes of the ground states.}
\label{tab:1}
\end{table}

\section{Conditional measurement of internal and motional states}
\label{sec:condmeas}

\begin{figure}
 \includegraphics[width=8.6cm]{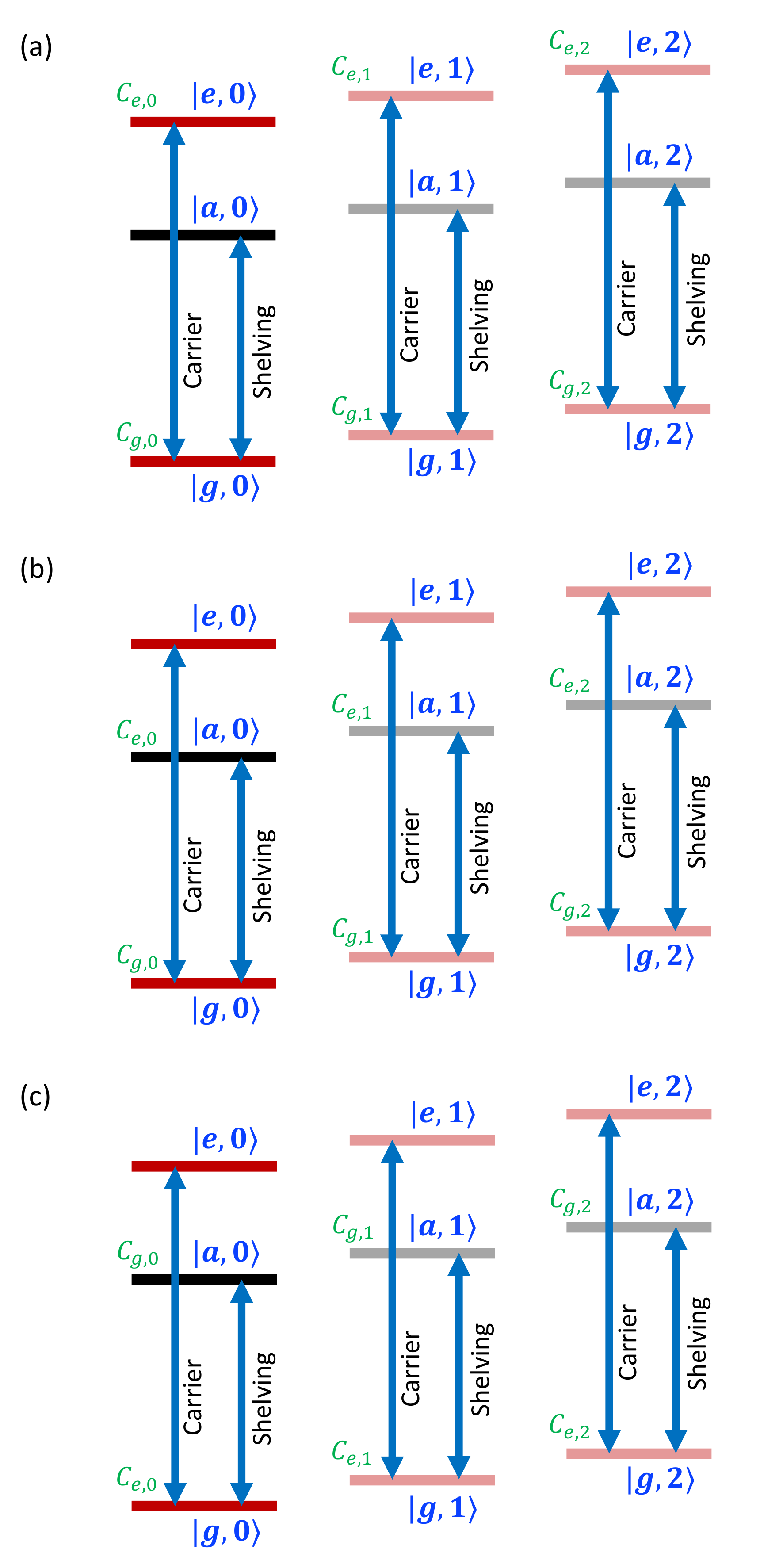}
 \caption {%
 (a) Level scheme and probability amplitudes associated with the levels 
 before a conditional measurement. 
 (b) Levels and probability amplitudes after conditioning the system to 
 $\ket{g}$ 
 [just after applying the first three pulses in 
 Fig.~\ref{conditional-sq}(b)]. 
 (c) Levels and probability amplitudes after conditioning the system to 
 $\ket{e}$
 [just after applying the first two pulses in 
 Fig.~\ref{conditional-sq}(c)].}
 \label{condg}    
\end{figure}

The levels used for the conditional-measurement scheme and 
the probability amplitudes associated with the levels 
are depicted in Fig.~\ref{condg}(a).
In this scheme, an auxiliary long-lived internal state
(a sublevel in the Zeeman manifold of the excited state; denoted as $\ket{a}$)
is used.
The levels in the internal two-level system $\{\ket{g},\ket{e}\}$
are connected by a carrier transition
(denoted as ``Carrier'' in Fig.~\ref{condg}), 
while the internal
ground state $\ket{g}$ and the auxiliary state $\ket{a}$ are connected 
by another carrier transition (denoted as ``Shelving'' in Fig.~\ref{condg}).
We assume that the motional states can be treated as a set of local vibrational 
modes.
This is an approximate picture, which is valid as long as
the sideband Rabi frequencies used for the analysis of the motional states
during the conditional measurements are much larger than the 
inter-site coupling rates of the motional states.

\begin{figure}
  \includegraphics[width=8.6cm]{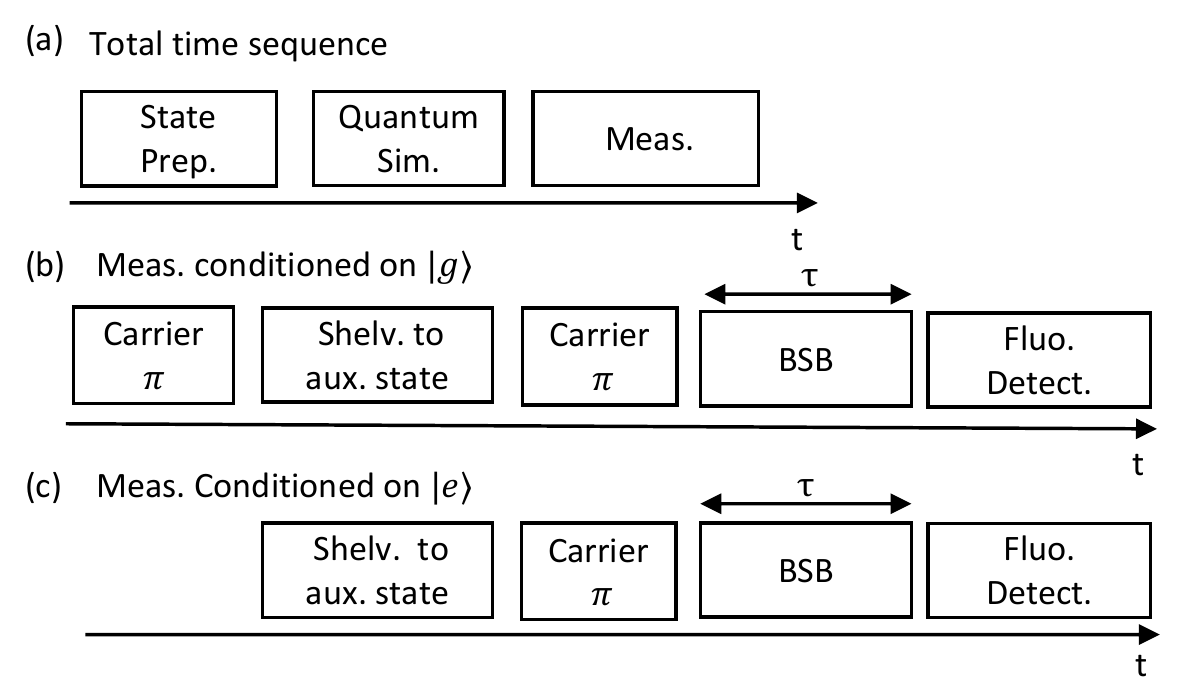}
  \caption{Time sequences for conditional measurements. 
(a) Total time sequence.
First, the initial state is prepared. Then a quantum simulation (an adiabatic transfer) and a conditional measurement are performed. 
(b) Time sequence for a measurement conditioned on $\ket{g}$. 
$\tau$ is the variable length of the BSB pulse, which is swept
to obtain the BSB oscillation signals.
(c) Time sequence for a measurement conditioned on $\ket{e}$. 
$\tau$ is a variable length similar to the one above.
}
  \label{conditional-sq}
\end{figure}

The time sequences for the conditional-measurement scheme are given in Fig.~\ref{conditional-sq}.
In the overall time sequence given in Fig.~\ref{conditional-sq}(a), 
an initial state is first prepared.
This involves cooling the system to the 
vibrational ground state and optical pumping to a particular Zeeman sublevel
in the internal ground-state manifold.
Then a quantum simulation (equivalent to the adiabatic transfer mentioned above) and a conditional measurement are performed. 

The conditional measurement is performed in two different ways: 
one is conditioned on $\ket{g}$ [Fig.~\ref{conditional-sq}(b)]
and the other on $\ket{e}$ [Fig.~\ref{conditional-sq}(c)].
We explain these two cases using a formulation, assuming
a chain of $N$ ions, where each ion has three internal states 
and an associated local vibrational mode.
The Coulomb couplings between different ion sites are ignored temporarily 
for simplicity.
The state of the $\it{i}$th ion is expressed as follows:

\begin{equation}
    \ket{\psi_i} = \sum_{x=g,e}\ket{\psi_x}_i,
\end{equation}
where 
\begin{equation}
   \left\{
   \begin{aligned}
    \ket{\psi_g}_i &=& \sum_{n=0}^m C_{g,n}\ket{ g,n}_i, \\
    \ket{\psi_e}_i &=& \sum_{n=0}^m C_{e,n}\ket{ e,n}_i.
   \end{aligned}
   \right.
\end{equation}

Here, $\ket{g,n}_{i}\equiv\ket{g}_{i}\ket{n}_{i}$ and $\ket{e,n}_{i}\equiv\ket{e}_{i}\ket{n}_{i}$, where
$n$ is the phonon number.
It is assumed that the maximum number of phonons in each vibrational mode is $m$. 
The coefficients $C_{g,n}$ and $C_{e,n}$ are probability amplitudes satisfying $\sum_{n=0}^m(|C_{g,n}|^2+|C_{e,n}|^2)=1$.
[See Fig.~\ref{condg}(a) for the initial arrangement of the probability amplitudes.]

As a preparation step 
for performing an analysis of the motional state
in the conditional-measurement scheme,
we transfer the population in either $\ket{g}$ or
$\ket{e}$ to a long-lived auxiliary state $\ket{a}$, depending on which state we want to analyze. 

To analyze the probabilities associated with $\ket{g}$, 
we use the time sequence given in Fig.~\ref{conditional-sq}(b).
We first apply a $\pi$ pulse resonant to the transition 
$\ket{g}\leftrightarrow\ket{e}$. 
This is followed by a $\pi$ pulse resonant to the $\ket{g}\leftrightarrow\ket{a}$ transition and a second $\pi$ pulse resonant to the $\ket{g}\leftrightarrow\ket{e}$ transition. These three pulses effectively shelve populations originally in $\ket{e}$ to $\ket{a}$, and keep those originally in $\ket{g}$ in the same state.
The state after the application of these pulses is represented as follows:
\begin{equation}
    \ket{\psi_i} = \sum_{x=g,a}\ket{\psi_x}_i,
\end{equation}
where
\begin{equation}
\label{eq6}
   \left\{
   \begin{aligned}
    \ket{\psi_g}_i = \sum_{n=0}^m C_{g,n}\ket{ g,n}_i,\\
    \ket{\psi_a}_i = \sum_{n=0}^m C_{e,n}\ket{ a,n}_i.
   \end{aligned}
   \right.
\end{equation}
[See Fig.~\ref{condg}(b) for the arrangement of the probability amplitudes
in this case.]

By applying a BSB pulse to this state,
whose variable length $\tau$ 
is swept from 0 to a certain time,
Rabi oscillations
between $\ket{g,n}\leftrightarrow\ket{e,n+1}$ can be observed. 
By performing Fourier analysis on the results, 
it is possible to deduce
the probabilities $|C_{g,n}|^2$ originally associated with $\ket{g}$.

To measure the probabilities associated with $\ket{e}$,
we use a slightly different time sequence, shown in Fig.~\ref{conditional-sq}(c).
In this case, 
the first $\pi$ pulse used in the previous case is omitted,
while the following sequence is used as-is. Before applying the BSB pulse, the state is similar to Eq.~(\ref{eq6}), where $C_{g,n}$ and $C_{e,n}$ are
replaced with each other.
[See Fig.~\ref{condg}(c) for the arrangement of the probability amplitudes
in this case.]
By applying a BSB pulse with a variable length $\tau$ and 
performing Fourier analysis on the result,
the probabilities $|C_{e,n}|^2$  originally associated with $\ket{e}$ can be obtained.

Collecting the results of these measurements, 
it is possible to evaluate the internal and motional states.
For example the probability for states with
$\it{k}$ polaritons is given as follows:
\begin{equation}
 P^k_\mathrm{pol} = 
  \begin{cases}
   |C_{g,k}|^2, & (k=0) \\
   |C_{g,k}|^2+|C_{e,k-1}|^2. & (k\ge1)
  \end{cases}
\end{equation}

We should mention the effective fraction of measurements
in this scheme.
In contrast to the conditional-measurement schemes
used previously \cite{Ivanov2009, Wang2010},
which involve a dedicated measurement of the internal state
in advance of the measurement of the motional state in each sequence,
our scheme uses shelving of the probability amplitudes that are 
of no interest to an auxiliary internal state.
This helps avoid unwanted effects of heating from
adjacent ion sites during fluorescence cycles,
and omits post-selections.
However, the effective amount of information acquired
is reduced in this scheme due to the fact that part of the probability amplitudes are
hidden in the auxiliary level.
In this sense this scheme is not superior to those methods that
use post-selections \cite{Ivanov2009, Wang2010}.
The overall loss of information on average is estimated to be 50\%.
For example, if we perform 100 measurement trials conditioned on $\ket{g}$ 
and the same number of trials conditioned on $\ket{e}$,
the average net amount of effective trials would be 100 in total, and
the respective ratio of the effective trials over the total number
may differ between $\ket{g}$ and $\ket{e}$,
depending on the internal-state population.
The effective fraction of 50\% does not depend on the number of ions involved.
Therefore, it can be asserted that this scheme is scalable with respect
to the number of ions in this sense.

This loss of information
manifests in the observed results
as a reduction in the contrasts in the BSB Rabi oscillation signals
and associated offsets corresponding to the populations hidden to
the long-lived auxiliary state.
(The populations in the auxiliary state cannot be distinguished
from those in the excited state in our case.)
The contrast in the results conditioned on $\ket{g}$ or $\ket{e}$
should be proportional to the population in that state
before the conditional-measurement sequence.
Therefore, a fitting process should be adapted to such signals
with different contrasts and offsets.
We will discuss this in Sec. \ref{sec:VIIa}.

\section{Experimental Procedure}
\label{sec:expproc}

Two  $^{40}$Ca$^+$ ions are trapped in a linear Paul trap. 
An rf voltage of 23 MHz is applied for radial confinement. 
An rf quadrupole electric field with a frequency of 23 MHz is applied
for radial confinement,
and a dc potential is applied for axial confinement. 
The secular frequencies are
$(\omega_x, \omega_y , \omega_z)/2\pi=(2.6, 2.4, 0.15)$  MHz. The distance between the two ions is 20 $\mu$m and the hopping rate of local phonons \cite{Porras2004,Deng2008} is 3.8 kHz. 
The ions are first cooled by Doppler cooling along all directions,
and then their
vibrational motion along the two radial directions ($x$ and $y$) is
reduced to the ground state by sideband cooling 
\cite{Wineland1998,Sasura2001}. 
Doppler cooling is realized with 397-nm ($S_{1/2}$--$P_{1/2}$) and 866-nm ($D_{3/2}$--$P_{3/2}$) laser beams 
and sideband cooling is realized with 729-nm ($S_{1/2}$--$D_{5/2}$) and 854-nm ($D_{5/2}$--$P_{3/2}$) laser beams. 
Sideband cooling is performed with a repetition of a pulse section 20 times,
in each of which 
729- and 854-nm pulses are simultaneously applied for 400 $\mu$s.
The ions are intermittently optically pumped to $S_{1/2}$ by a 397-nm beam with $\sigma^-$ polarization before, between and after the pulse sections 
for sideband cooling.
The final vibrational quantum numbers along the radial directions
after sideband cooling are 
$\{\bar{n}_x,\bar{n}_y\}\sim\{0.2,0.1\}$.

The $y$ radial direction in the ion crystal is used for the motional degree
of interest in this work.
There are two collective modes in this direction, the center-of-mass (COM) and rocking modes, which are separated from each other in frequency by a spacing similar to the hopping rate.  
The heating rate for the $y$ direction is estimated to
be $\sim$5 quanta/s \cite{PhysRevLett.124.200501}.

\begin{figure}
 \includegraphics[width=8.6cm]{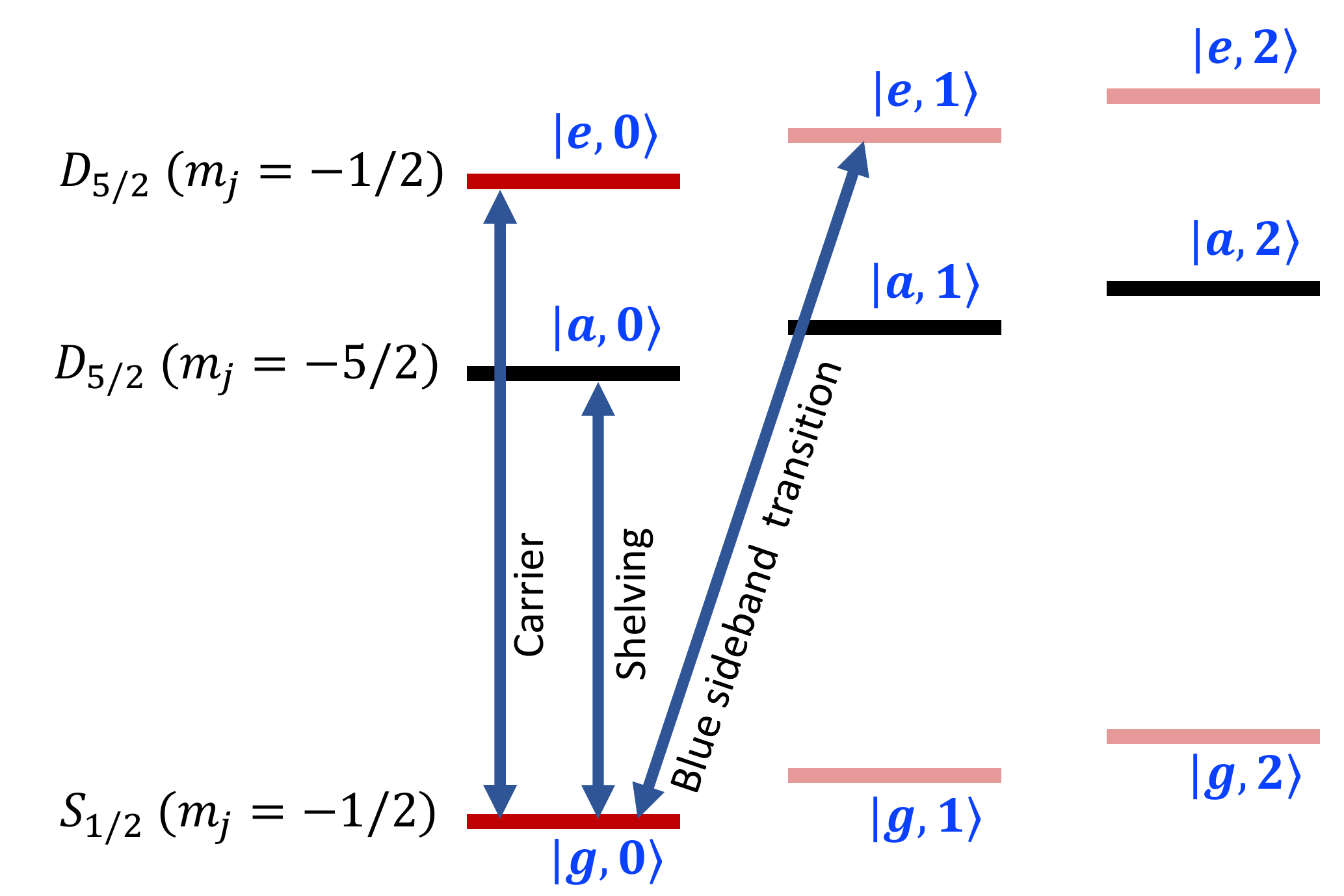}
 \caption { Level scheme for a trapped $^{40}$Ca$^+$ with internal ground ($\ket{g,n}$), excited ($\ket{e,n}$) and auxiliary ($\ket{a,n}$) states, 
where $n$ represents the motional quantum number in a local vibrational mode.}
     \label{level}
\end{figure}

Excitation beams at 729 nm for $S_{1/2}$--$D_{5/2}$ are used to induce JC coupling, as well as to prepare and analyze the internal and motional states. 
The level scheme of $^{40}$Ca$^+$ relevant for the manipulation of
the internal and motional states 
is shown in Fig.~\ref{level}. 
(In this figure the motional state is assumed to be that of 
a local vibrational mode in view of applying this to the measurement
scheme.)

The excitation beams at 729 nm are individually applied to each of the two
ions.
The relative intensities of the beams are adjusted to optimize equal illumination of the two ions, by balancing the maximum Rabi frequencies between the two ions. 

The fluctuations in the radial secular frequencies are reduced by introducing a feedback-control system for the rf amplitude \cite{Johnson2016}. 

The internal state of the two ions is determined by illuminating them with lasers at 397 and 866 nm and by detecting fluorescence photons with a photomultiplier tube or an electron multiplying charge-coupled-device camera. 

In the quantum simulation of the JCH model, first the system is prepared
in the atomic MI state
($\ket{\psi_\mathrm{MI}}=\ket{e_1}\ket{e_2}\ket{0}_1\ket{0}_2$)
by applying a carrier $\pi$ pulse at 729 nm
after sideband cooling.
Then an adiabatic transfer is performed, with a duration
of typically 960 $\mu$s. 
The duration is determined to satisfy the adiabaticity in the internal states,
while the adiabaticity in the motional state is not fully satisfied
(this will be discussed in Sec. \ref{sec:VIIc}).
The internal state during this adiabatic transfer
can be detected by truncating it and by illuminating the ions with the
lasers at 397 and 866 nm.
A conditional measurement, as explained above, involving
the excitation of the BSB transition,
can also be performed by truncating the adiabatic transfer.

\section{Measurement of internal-state populations}
\label{sec:measint}
\begin{figure}
     \includegraphics[width=8.6cm]{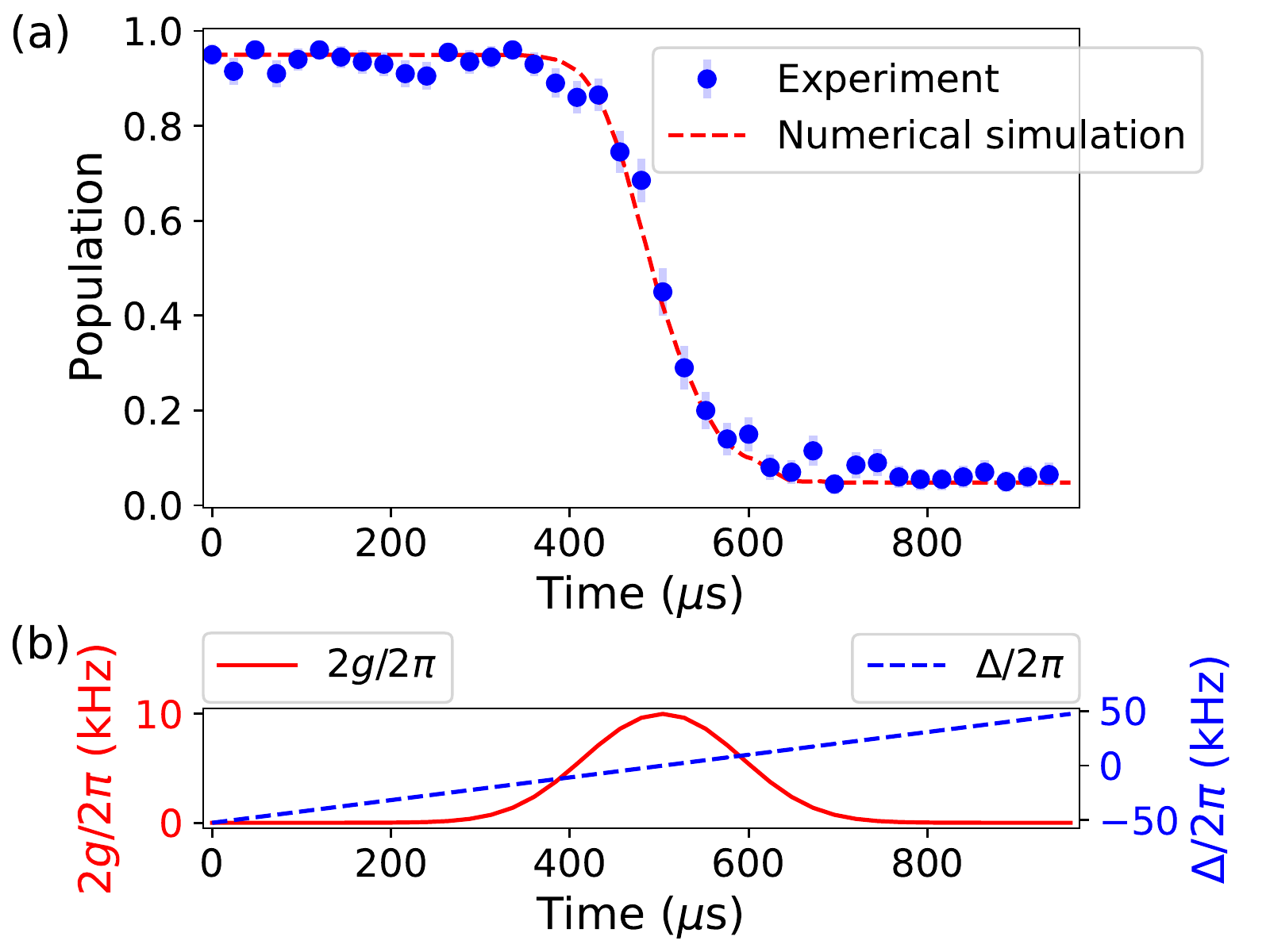}
     \caption { (a) Experimentally observed adiabatic-transfer curve 
(blue circles  and error bars) 
and numerically simulated results (red dashed curve). 
Each experimental point is the average of 100 experiments, and the error is calculated by assuming a binomial distribution.
(b) Time dependence of the parameters for the optical pulse 
assumed in the numerical simulation in (a): 
$g$ (JC-coupling constant) divided by $\pi$ (red solid curve) and $\Delta$ (detuning) divided by $2\pi$ (blue dashed curve) for the numerical simulation.}
     \label{rap}
\end{figure}

The results of internal state measurements from 0 $\mu$s to 960 $\mu$s are shown in Fig.~\ref{rap}. 
The measured population (blue circles with error bars)
and a numerical simulation for the population
 (red dashed curve) are
shown in Fig.~\ref{rap}(a).
The time dependence of the parameters
used in the numerical simulation in Fig.~\ref{rap}(a) is plotted 
in Fig.~\ref{rap}(b). 
The conditions for the numerical simulation are summarized as follows.
The hopping rate $\kappa/2\pi$ is 3.8kHz.
$2g/2\pi$ is swept in a Gaussian form with a maximum value of 10.0 kHz.
$\Delta/2\pi$ is varied from $-$50 kHz to 50 kHz during a time of 960 $\mu s$.
The Lindblad master equation \cite{Manzano2020} is used for the numerical simulation, 
in which transverse relaxation for the BSB transition,
for example due to fluctuations of the sideband resonance frequencies, 
is taken into account by using $\gamma_\mathrm{T}=2\pi\times0.19$ kHz as the rate.
The initial populations in the internal states are set to be 0.95 in $\ket{e}$ and 0.05 in $\ket{g}$
\footnote{%
The cause of the reduction by 5 \% in the population of $\ket{e}$ can be explained as being due to the incomplete preparation of the atomic MI state by the carrier $\pi$ pulse. 
We use the individual illumination of each ion with a dedicated beam at 729 nm.
Each beam is passed through a dedicated polarization-maintaining optical fiber
to assure path stability. However, this fiber may produce a drift in the optical phases.
Fluctuations in interference between the beams caused by this drift of optical phases, as well as those
in the pointing of the beams, 
lead to relatively large intensity fluctuations at the positions of the ions and affect the fidelity of rotations
even for the carrier transition.%
}.

The greater than zero final population can be explained mainly by the effect of
transverse relaxation.
In our estimation,  the effect of diabaticity is not significant
for this particular case.

The transfer of population
from $\ket{e}$ to $\ket{g}$ observed in Fig.~\ref{rap}(a)
can be considered to be evidence that
a transition from the atomic MI state to the phonon SF state
has occurred \cite{Toyo2013}.

\section{Conditional measurements and evaluation of variances}
\label{sec:meascond}
\subsection{Evaluation of populations from fitting}
\label{sec:VIIa}
In performing the conditional measurements to evaluate the variances, 
we applied the two different time sequences given 
in Fig.~\ref{conditional-sq}(b) and (c)
in separate experimental runs,
and the BSB Rabi oscillations in both cases were measured. 
By analyzing the Rabi oscillation results in both cases 
via fitting with model functions, and by combining the acquired information,
the state populations were inferred and the variances were then evaluated.

The atomic variance can be estimated if the internal state population is
known.
For the particular conditional measurements in this work, 
we used the first points of the BSB Rabi oscillations,
for which no BSB pulse of a finite length is applied, 
to infer the population in the internal states.
These are equivalent to the internal-state populations
just before the application of the BSB pulses
in Fig.~\ref{conditional-sq}(b) and (c).
Those populations are denoted as $P_g$ and $P_e$ for the ground and excited 
states, respectively.
Here, we assume that the system is symmetric with respect to the
permutation of the two ions, and hence that the populations for
the two ions are identical to each other.

For the estimation of the phonon and total variances, especially for 
the latter,
the information obtained from conditional measurements
is fully exploited.
The obtained BSB Rabi oscillation signals are
analyzed by fitting the obtained results with model functions 
derived from those used in standard analyses on sideband Rabi
oscillations \cite{Meekhof1996}:

\begin{eqnarray*}
    \Pi_{e}(t) &=& \frac{1}{2}a\left( 1 + \sum_{n=0}^{m} \tilde{P}_n \cos^2 (\Omega_{n,n+1} t/2)e^{-\gamma_\mathrm{R} t} \right)\\
&&+(1-a).
\end{eqnarray*}
Here, $\Pi_{e}(t)$ is the time-dependent excited-state population
during the BSB excitation,
$m$ is the maximum Fock-state quantum number assumed,
$\tilde{P}_n$ is the population in each motional Fock state
before the BSB excitation,
$\Omega_{n,n+1}\equiv\eta\Omega_0\sqrt{n+1}$ 
is the BSB Rabi frequency
($\eta$ is the Lamb Dicke factor and $\Omega_0$ is the resonant Rabi frequency
for an ion without motion; 
the terms of higher orders in $\eta$ are ignored here),
and $\gamma_\mathrm{R}$ is the common relaxation rate reflecting decoherence
and dephasing processes in sideband Rabi oscillations
\footnote{
In principle, the previously quoted transverse-relaxation rate $\gamma_\mathrm{T}$ in the Lindblad master equation and
this $\gamma_\mathrm{R}$ should have a certain relation.
If the latter is assumed to be determined purely from the former,  $\gamma_\mathrm{R}\sim0.25\times\gamma_\mathrm{T}$
is expected, which is confirmed by numerical simulations.
In that case, $\gamma_\mathrm{R}\sim0.05$ kHz is expected if $\gamma_\mathrm{T}\sim2\pi\times0.19$ kHz is satisfied.
On the other hand, the actual value of $\gamma_\mathrm{R}$ determined from the chi-square minimization, as explained in Sec. \ref{sec:VIIa},
is $2\pi\times1.1$ kHz, which is more than an order of magnitude larger than the above-mentioned value.
This discrepancy may be explained by the concrete nature of the fluctuations.
If an adiabatic-transfer process like that in Fig.~\ref{rap}(a) is used and
the typical timescale for the fluctuations is much longer than the pulse duration,
the adiabatic transfer process is not affected significantly by the fluctuations thanks to their robustness against parameter variations.
This may be why the assumed relaxation rate that matches the experimental results
is much smaller than that for the numerical simulation results shown in Fig.~\ref{rap}(a).
}.
$a$ ($0\le{a}\le1$) is a factor that
takes into account the reduction of contrasts and the associated offsets
as discussed above.
Here, $a$ is determined from the estimated internal-state populations as explained in the last paragraph. 

In the actual fitting process, 
the maximum Fock-state quantum number $m$ is set to 2 
\footnote{
The factors that may make this assumption invalid are
the residual phonon number after sideband cooling 
(estimated to be $\sim0.2$ in the present case),
heating during adiabatic transfer ($<\sim5\times10^{-3}$ quanta),
and off-resonance excitation of the carrier transition
by the red-sideband optical pulse for the adiabatic transfer
($<0.03$).
Here, we ignore their effects.
}.
The base Rabi frequency $\Omega_{0,1}$ and
$(\tilde{P}_0,\tilde{P}_1,\tilde{P}_2)$ ($\tilde{P}_0+\tilde{P}_1+\tilde{P}_2=1$)
for each result, as well as 
the overall relaxation rate $\gamma_\mathrm{R}$, 
are determined iteratively.
First, we use empirically known fixed values of $(\tilde{P}_0,\tilde{P}_1,\tilde{P}_2)$ for each
result,
and perform a fitting of all the results 
with the least-squares method
to determine
$\Omega_{0,1}$ for each result and $\gamma_\mathrm{R}$
(the latter is determined by finding the value that gives
the minimum value for the sum of the chi-squares in fitting of all the results).
We then treat these values as fixed values and repeat the fitting process,
thereby determining the refined values for $(\tilde{P}_0,\tilde{P}_1,\tilde{P}_2)$.
This whole process is repeated multiple times to obtain the final values 
for $(\tilde{P}_0,\tilde{P}_1,\tilde{P}_2)$.
By multiplying $a$ by these values, the actual populations
in the combined internal and motional state basis states are determined.
For example, in the case for conditioning on $\ket{g}$ or $\ket{e}$, 
we obtain
$(P_{g,0},P_{g,1},P_{g,2})$ or 
$(P_{e,0},P_{e,1},P_{e,2})$, respectively,
which is equal to $(a\tilde{P}_0,a\tilde{P}_1,a\tilde{P}_2)$.
Here, 
$P_{x,n}\equiv|C_{x,n}|^2$ ($x=g,e$) is the population in each
combined internal and motional basis state $\ket{x,n}$. 
We also evaluate the error for each of the populations,
which is calculated as that
propagating from the parameter errors
in the fitting process.

\subsection{Evaluation of variances}
\label{sec:VIIb}
The variances defined in Sec. \ref{sec:misf},
including the total, atomic and phonon variances,
 are obtained from the combined information
acquired in both the cases conditioned on $\ket{g}$ and $\ket{e}$. 
In this subsection we present the method to calculate the variances 
using populations based on the method
in the previous subsection.
As mentioned above, we assume that the system is symmetric with respect to the
permutation of the two ions, and omit the subscript
$j=1,2$ representing the index for the ion number 
in the following.
The atomic variance is obtained as follows:
\begin{eqnarray*}
\Delta \hat{N}_{a}^2&=& \mathrm{tr}(\rho \hat{N}_{a}^2) - 
\mathrm{tr}(\rho\hat{N}_{a})^2\\
&=&P_{e}-P_{e}^2, 
\end{eqnarray*}
where $\rho$ is the density operator for the total system
and $P_{e}=\bra{e}\rho\ket{e}$ (this is equal to $\sum_{n=0}^2P_{e,n}$
when the phonon number $n$ is limited to $0\le{}n\le2$).
The phonon variance is obtained as follows:
\begin{equation*}
 \Delta \hat{N}_p^2=\sum_{n=0}^2n^2P_n-\left(\sum_{n=0}^2nP_n\right)^2,
\end{equation*}
where $P_n\equiv{}P_{g,n}+P_{e,n}$ ($n=0,1,2$).
The total variance can be expressed as follows:
\begin{eqnarray*}
\Delta\hat{N}^2&=&
\langle(\hat{N}_a+\hat{N}_p)^2\rangle-\langle\hat{N}_a+\hat{N}_p\rangle^2\\
&=&(\langle\hat{N}_a^2\rangle-\langle\hat{N}_a\rangle^2)
+(\langle\hat{N}_p^2\rangle-\langle\hat{N}_p\rangle^2)\\
&&+2(\langle\hat{N}_a\hat{N}_p\rangle-\langle\hat{N}_a\rangle\langle\hat{N}_p\rangle)\\
&=&\Delta\hat{N}_a^2+\Delta\hat{N}_p^2+2\mathrm{Cov}(\hat{N}_a,\hat{N}_p),
\end{eqnarray*}
where 
\begin{eqnarray*}
\mathrm{Cov}(\hat{N}_a,\hat{N}_p)&\equiv&
\langle\hat{N}_a\hat{N}_p\rangle-\langle\hat{N}_a\rangle\langle\hat{N}_p\rangle\\
&=&\sum_{n=0}^2nP_{e,n}-P_e\sum_{n=0}^2nP_n
\end{eqnarray*}
is the covariance between the atomic excitation number and the phonon number.
We also evaluate the error for each of the variances,
which is calculated as that
propagating from those of the evaluated values for the populations
\footnote{Here, for simplicity, we assume that the errors are symmetric 
for the positive and negative directions, 
and nonlinear dependences of the deviations around the estimated values
are ignored. This could lead to the appearance of unphysical values 
for the bounds of the confidence intervals: 
the negative values in Fig.~\ref{sbvar}(a) and (b) at 96 and 192 $\mu$s
are considered to be due to this.}.

\subsection{Experimental results}
\label{sec:VIIc}
\begin{figure}
     \hspace*{-1cm}
     \includegraphics[width=18cm]{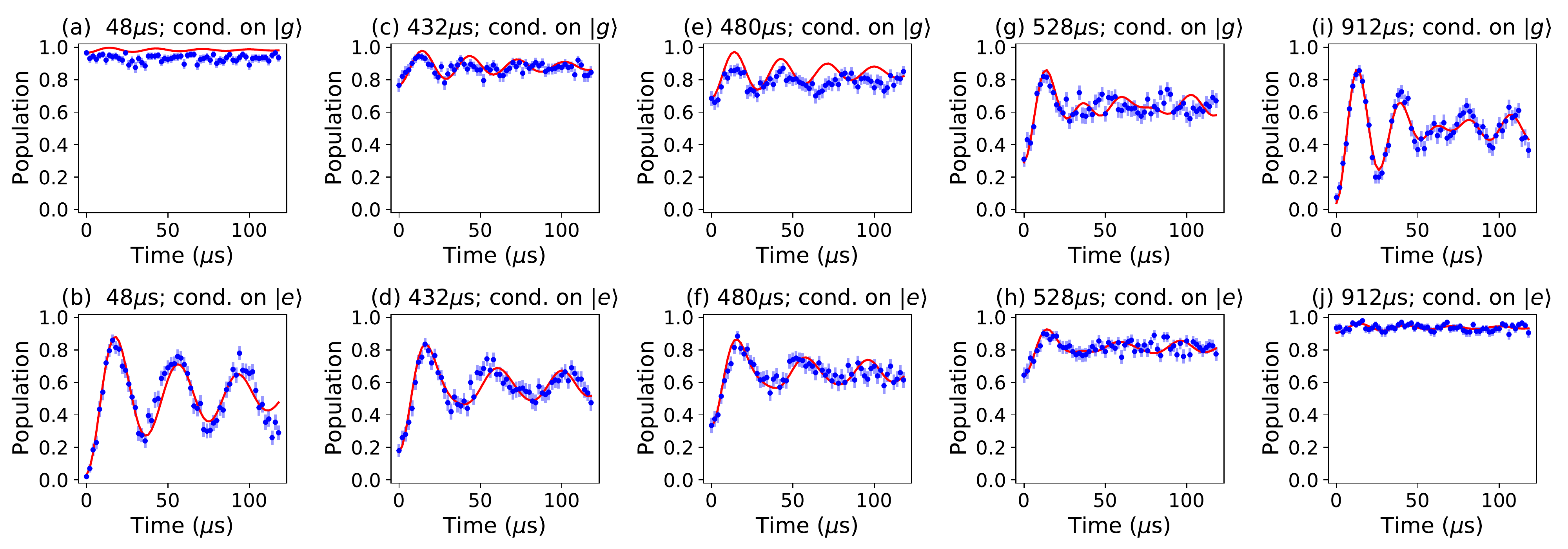}
     \caption {BSB Rabi oscillations (blue points) obtained by truncating the quantum simulation (adiabatic transfer) and by illuminating using a square pulse with a variable duration. 
Here, only certain values for the duration of the quantum simulation are picked up.
Each experimental point is the average of 100 experiments, and the error is calculated by assuming a binomial distribution.
The results for the fitting of the Rabi-oscillation signals with model functions are also shown (red curves).  The durations for the quantum simulation (adiabatic transfer) are 
(a--b) 48, 
(c--d) 432, 
(e--f) 480, 
(g--h) 528 and
(i--j) 912 $\mu$s from the first column to the fifth.
The results in the upper row [(a), (c), (e), (g) and (i)] are taken
from measurements conditioned on $\ket{g}$,
and those in the lower row [(b), (d), (f), (h) and (j)]
on $\ket{e}$.}
     \label{fit}
\end{figure}

\begin{figure}
\begin{center}
\hspace*{-1cm}
 \includegraphics[width=18cm]{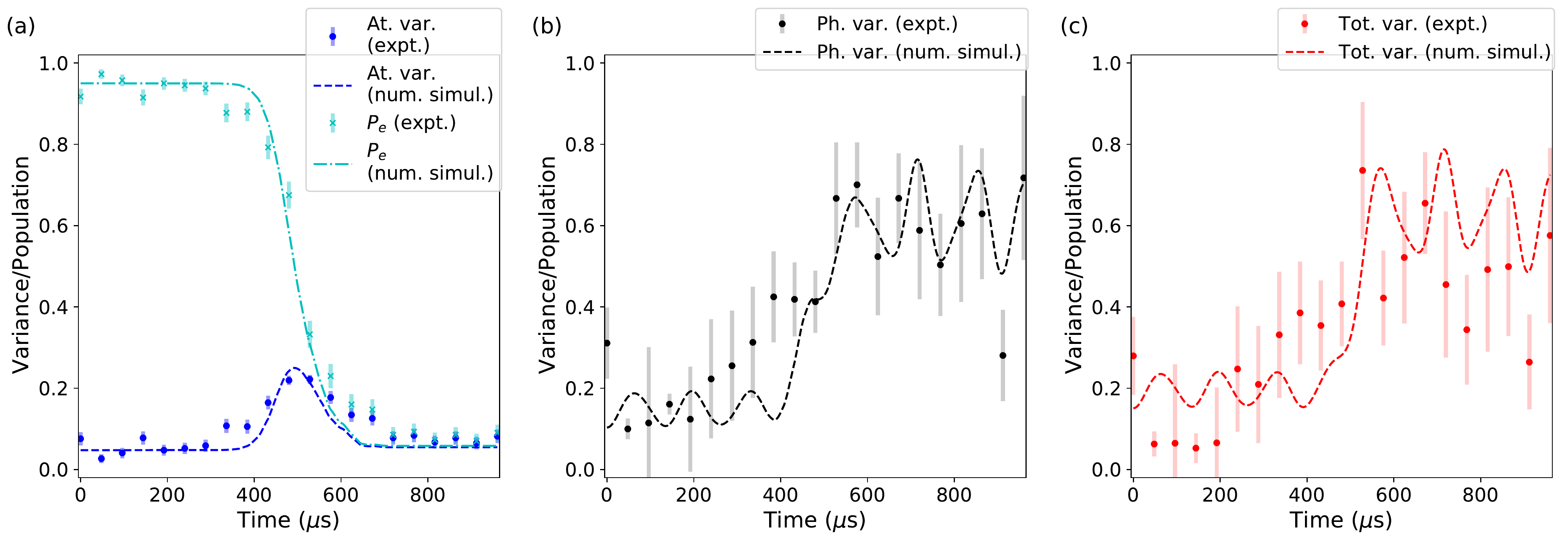}
\end{center}
 \caption {
(a) Atomic variance, excited-state population $P_e$, (b) phonon variance and 
(c) total variance, which are
obtained from an analysis
involving the fitting of experimental results (points with error bars)
and numerical simulation (dashed/dash-dot curves).
The irregularly high experimental values ($\sim0.3$) at 0 $\mu$s in (b) and (c) are considered to be
artifacts due to technical imperfections.
}
 \label{sbvar} 
\end{figure}

The experimental Rabi oscillation results measured after truncating
the adiabatic transfer at certain selected points in time
(48,
432,
480,
528 and
912 $\mu$s),
as well as the results of fitting performed based on the above-mentioned
procedure,
are shown in Fig.~\ref{fit}.
Values for the variances and internal-state population estimated
by processing the experimental Rabi oscillation results,
as well as those obtained in a numerical simulation,
are shown in Fig.~\ref{sbvar}. 
The atomic variance 
[blue circles with error bars and dashed curve in Fig.~\ref{sbvar}(a)]
shows a peak in the center.
In this region, the excitations of the system take the form of polaritons, each of which is the superposition of an internal excitation and a phonon. In this case, the internal-excitation number fluctuates, resulting in the appearance of a peak in the atomic variance.
The phonon number variance 
[black circles with error bars and dashed curve in Fig.~\ref{sbvar}(b)]
increases around the center of the sweep.
This indicates the emergence of the phonon SF state. 
The total variance 
[red circles with error bars and dashed curve in Fig.~\ref{sbvar}(c)]
follows a similar trend.

In the numerically simulated results in Fig.~\ref{sbvar}(b--c), 
at the region after
500 $\mu$s, oscillatory behavior with a period of $\sim$150 $\mu$s
is seen in both the phonon and total variances.  
This feature may have arisen from 
the diabaticity in the motional states.  Since the total sweep time of 960 $\mu$s is not sufficiently long
compared with the hopping time constant (defined as the inverse of the hopping
rate, $\sim$260 $\mu$s), the system does not completely adiabatically
follow the ground states and additional phonon excitations may be
generated.  
This effect may be relaxed by choosing a longer sweep time.
We suppose that the relatively large fluctuations in the experimental
phonon and total variances [Fig.~\ref{sbvar}(b--c)] observed in the region
after 500 $\mu$s have an origin related to this effect. 

The internal-state population 
[cyan crosses with error bars and dash-dot curve in Fig.~\ref{sbvar}(a)]
shows a relatively smooth and solid transition.
This indicates that adiabaticity of the internal states is satisfied. 
This experiment confirms that 
the method for conditional measurements described in this work,
which is capable of measuring the internal and motional states simultaneously
without mutual interference,
is sufficient for investigating the quantum phase transitions in the current system.



\section{Discussion}
\label{sec:discussion}

The study of the ground-state properties in a JCH system,
which can be dealt with by the conditional-measurement method in this work,
may provide a basis for dynamical studies of the system as well as many-body 
quantum dynamics. 
The study of correlated particles such as polaritons may lead us toward more profound
quantum mechanical properties or phenomena.  

The conditional-measurement method described in this work can be applied to larger numbers of ions in a straightforward manner.
By increasing the number of ions, a steep transition from one quantum phase to another is expected \cite{Islam2011}. In a previous study using a Bose-particle system, a sudden change of quantum phases against parameter changes was observed \cite{greiner2002}. The JCH system is expected to show a similar steep variation of quantum phases against parameter changes if an increased number of ions is used.

The fitting model functions used in this article 
assume that sideband Rabi oscillations
are bound within the state manifold for each ion, and mutual interactions
between ions are not considered.
Although this assumption simplifies the analysis, it can 
lead to discrepancy between the experimental and fitted results.
This discrepancy could be overcome if the full JCH model is used
to obtain the fitting models in numerical simulations.
However, this way of analysis is not considered to be scalable,
since the numerical simulation of the full JCH model becomes
increasingly difficult for larger numbers of ion sites and
is effectively intractable
\footnote{
In our estimation, numerical simulation using the full JCH model
with more than 17 ion sites and the commensurate filling
of one excitation per ion site requires memory that
can hold a number of basis states as large as that in a 40-spin system
and cannot be easily solved with existing classical computers.}.
In addition, fitting necessitates setting suitable initial values,
which may allow arbitrariness to enter into the analysis process. 

Another option that can overcome the shortcomings of using fitting
for the analysis 
would be the use of discrete Fourier transforms.
By not relying on fitting or
numerical simulations of physical models to deduce phonon distributions,
this method of analysis may be
applied to larger numbers of ion sites with less difficulties and arbitrariness.

\section{Conclusions}
\label{sec:conclusions}

The system of polaritons in the JCH model that undergoes quantum phase transitions is studied with a trapped-ion system. 
An MI-to-SF transition occurs via polaritonic intermediate states. 
The polariton states are measured by conditional measurements in which the internal state and motional state can be measured simultaneously.

\section*{Acknowledgments}
This article was supported by MEXT Quantum Leap Flagship Program (MEXT Q-LEAP) Grant Number JPMXS0118067477.

\subsection*{}
S.M. and R.O. contributed equally to this article.

\bibliography{final.bbl}

\begin{thebibliography}{48}%
\makeatletter
\providecommand \@ifxundefined [1]{%
 \@ifx{#1\undefined}
}%
\providecommand \@ifnum [1]{%
 \ifnum #1\expandafter \@firstoftwo
 \else \expandafter \@secondoftwo
 \fi
}%
\providecommand \@ifx [1]{%
 \ifx #1\expandafter \@firstoftwo
 \else \expandafter \@secondoftwo
 \fi
}%
\providecommand \natexlab [1]{#1}%
\providecommand \enquote  [1]{``#1''}%
\providecommand \bibnamefont  [1]{#1}%
\providecommand \bibfnamefont [1]{#1}%
\providecommand \citenamefont [1]{#1}%
\providecommand \href@noop [0]{\@secondoftwo}%
\providecommand \href [0]{\begingroup \@sanitize@url \@href}%
\providecommand \@href[1]{\@@startlink{#1}\@@href}%
\providecommand \@@href[1]{\endgroup#1\@@endlink}%
\providecommand \@sanitize@url [0]{\catcode `\\12\catcode `\$12\catcode
  `\&12\catcode `\#12\catcode `\^12\catcode `\_12\catcode `\%12\relax}%
\providecommand \@@startlink[1]{}%
\providecommand \@@endlink[0]{}%
\providecommand \url  [0]{\begingroup\@sanitize@url \@url }%
\providecommand \@url [1]{\endgroup\@href {#1}{\urlprefix }}%
\providecommand \urlprefix  [0]{URL }%
\providecommand \Eprint [0]{\href }%
\providecommand \doibase [0]{https://doi.org/}%
\providecommand \selectlanguage [0]{\@gobble}%
\providecommand \bibinfo  [0]{\@secondoftwo}%
\providecommand \bibfield  [0]{\@secondoftwo}%
\providecommand \translation [1]{[#1]}%
\providecommand \BibitemOpen [0]{}%
\providecommand \bibitemStop [0]{}%
\providecommand \bibitemNoStop [0]{.\EOS\space}%
\providecommand \EOS [0]{\spacefactor3000\relax}%
\providecommand \BibitemShut  [1]{\csname bibitem#1\endcsname}%
\let\auto@bib@innerbib\@empty
\bibitem [{\citenamefont {Feynman}(1982)}]{feynman1982}%
  \BibitemOpen
  \bibfield  {author} {\bibinfo {author} {\bibfnamefont {R.~P.}\ \bibnamefont
  {Feynman}},\ }\bibfield  {title} {\bibinfo {title} {Simulating physics with
  computers},\ }\href@noop {} {\bibfield  {journal} {\bibinfo  {journal} {Int.
  J. Theor. Phys.}\ }\textbf {\bibinfo {volume} {21}},\ \bibinfo {pages} {467}
  (\bibinfo {year} {1982})}\BibitemShut {NoStop}%
\bibitem [{\citenamefont {Blatt}\ and\ \citenamefont {Roos}(2012)}]{Blatt2012}%
  \BibitemOpen
  \bibfield  {author} {\bibinfo {author} {\bibfnamefont {R.}~\bibnamefont
  {Blatt}}\ and\ \bibinfo {author} {\bibfnamefont {C.~F.}\ \bibnamefont
  {Roos}},\ }\bibfield  {title} {\bibinfo {title} {Quantum simulations with
  trapped ions},\ }\href@noop {} {\bibfield  {journal} {\bibinfo  {journal}
  {Nat. Phys.}\ }\textbf {\bibinfo {volume} {8}},\ \bibinfo {pages} {277}
  (\bibinfo {year} {2012})}\BibitemShut {NoStop}%
\bibitem [{\citenamefont {Monroe}\ \emph {et~al.}(2021)\citenamefont {Monroe},
  \citenamefont {Campbell}, \citenamefont {Duan}, \citenamefont {Gong},
  \citenamefont {Gorshkov}, \citenamefont {Hess}, \citenamefont {Islam},
  \citenamefont {Kim}, \citenamefont {Linke}, \citenamefont {Pagano},
  \citenamefont {Richerme}, \citenamefont {Senko},\ and\ \citenamefont
  {Yao}}]{Monroe2019}%
  \BibitemOpen
  \bibfield  {author} {\bibinfo {author} {\bibfnamefont {C.}~\bibnamefont
  {Monroe}}, \bibinfo {author} {\bibfnamefont {W.~C.}\ \bibnamefont
  {Campbell}}, \bibinfo {author} {\bibfnamefont {L.-M.}\ \bibnamefont {Duan}},
  \bibinfo {author} {\bibfnamefont {Z.-X.}\ \bibnamefont {Gong}}, \bibinfo
  {author} {\bibfnamefont {A.~V.}\ \bibnamefont {Gorshkov}}, \bibinfo {author}
  {\bibfnamefont {P.}~\bibnamefont {Hess}}, \bibinfo {author} {\bibfnamefont
  {R.}~\bibnamefont {Islam}}, \bibinfo {author} {\bibfnamefont
  {K.}~\bibnamefont {Kim}}, \bibinfo {author} {\bibfnamefont {N.~M.}\
  \bibnamefont {Linke}}, \bibinfo {author} {\bibfnamefont {G.}~\bibnamefont
  {Pagano}}, \bibinfo {author} {\bibfnamefont {P.}~\bibnamefont {Richerme}},
  \bibinfo {author} {\bibfnamefont {C.}~\bibnamefont {Senko}},\ and\ \bibinfo
  {author} {\bibfnamefont {N.}~\bibnamefont {Yao}},\ }\bibfield  {title}
  {\bibinfo {title} {Programmable quantum simulations of spin systems with
  trapped ions},\ }\href@noop {} {\bibfield  {journal} {\bibinfo  {journal}
  {Rev. Mod. Phys.}\ }\textbf {\bibinfo {volume} {93}},\ \bibinfo {pages}
  {025001} (\bibinfo {year} {2021})}\BibitemShut {NoStop}%
\bibitem [{\citenamefont {Jaynes}\ and\ \citenamefont
  {Cummings}(1963)}]{Jaynes1963}%
  \BibitemOpen
  \bibfield  {author} {\bibinfo {author} {\bibfnamefont {E.~T.}\ \bibnamefont
  {Jaynes}}\ and\ \bibinfo {author} {\bibfnamefont {F.~W.}\ \bibnamefont
  {Cummings}},\ }\bibfield  {title} {\bibinfo {title} {Comparison of quantum
  and semiclassical radiation theories with application to the beam maser},\
  }\href@noop {} {\bibfield  {journal} {\bibinfo  {journal} {Proc. IEEE}\
  }\textbf {\bibinfo {volume} {51}},\ \bibinfo {pages} {89} (\bibinfo {year}
  {1963})}\BibitemShut {NoStop}%
\bibitem [{\citenamefont {Greentree}\ \emph {et~al.}(2006)\citenamefont
  {Greentree}, \citenamefont {Tahan}, \citenamefont {Cole},\ and\ \citenamefont
  {Hollenberg}}]{Greentree2006}%
  \BibitemOpen
  \bibfield  {author} {\bibinfo {author} {\bibfnamefont {A.~D.}\ \bibnamefont
  {Greentree}}, \bibinfo {author} {\bibfnamefont {C.}~\bibnamefont {Tahan}},
  \bibinfo {author} {\bibfnamefont {J.~H.}\ \bibnamefont {Cole}},\ and\
  \bibinfo {author} {\bibfnamefont {L.~C.~L.}\ \bibnamefont {Hollenberg}},\
  }\bibfield  {title} {\bibinfo {title} {Quantum phase transitions of light},\
  }\href@noop {} {\bibfield  {journal} {\bibinfo  {journal} {Nat. Phys.}\
  }\textbf {\bibinfo {volume} {2}},\ \bibinfo {pages} {856} (\bibinfo {year}
  {2006})}\BibitemShut {NoStop}%
\bibitem [{\citenamefont {Hartmann}\ \emph {et~al.}(2006)\citenamefont
  {Hartmann}, \citenamefont {Brandao},\ and\ \citenamefont
  {Pleino}}]{hartmann2006}%
  \BibitemOpen
  \bibfield  {author} {\bibinfo {author} {\bibfnamefont {M.~J.}\ \bibnamefont
  {Hartmann}}, \bibinfo {author} {\bibfnamefont {F.}~\bibnamefont {Brandao}},\
  and\ \bibinfo {author} {\bibfnamefont {M.~B.}\ \bibnamefont {Pleino}},\
  }\bibfield  {title} {\bibinfo {title} {Strongly interacting polaritons in
  coupled arrays of cavities},\ }\href@noop {} {\bibfield  {journal} {\bibinfo
  {journal} {Nat. Phys.}\ }\textbf {\bibinfo {volume} {2}},\ \bibinfo {pages}
  {849} (\bibinfo {year} {2006})}\BibitemShut {NoStop}%
\bibitem [{\citenamefont {Angelakis}\ \emph {et~al.}(2007)\citenamefont
  {Angelakis}, \citenamefont {Santos},\ and\ \citenamefont
  {Bose}}]{Angelakis2007}%
  \BibitemOpen
  \bibfield  {author} {\bibinfo {author} {\bibfnamefont {D.~G.}\ \bibnamefont
  {Angelakis}}, \bibinfo {author} {\bibfnamefont {M.~F.}\ \bibnamefont
  {Santos}},\ and\ \bibinfo {author} {\bibfnamefont {S.}~\bibnamefont {Bose}},\
  }\bibfield  {title} {\bibinfo {title} {Photon-blockade-induced {M}ott
  transitions and {XY} spin models in coupled cavity arrays},\ }\href@noop {}
  {\bibfield  {journal} {\bibinfo  {journal} {Phys. Rev. A}\ }\textbf {\bibinfo
  {volume} {76}},\ \bibinfo {pages} {031805} (\bibinfo {year}
  {2007})}\BibitemShut {NoStop}%
\bibitem [{\citenamefont {Hartmann}\ \emph {et~al.}(2008)\citenamefont
  {Hartmann}, \citenamefont {Brandao},\ and\ \citenamefont
  {Plenio}}]{Hartmann2008}%
  \BibitemOpen
  \bibfield  {author} {\bibinfo {author} {\bibfnamefont {M.~J.}\ \bibnamefont
  {Hartmann}}, \bibinfo {author} {\bibfnamefont {F.}~\bibnamefont {Brandao}},\
  and\ \bibinfo {author} {\bibfnamefont {M.~B.}\ \bibnamefont {Plenio}},\
  }\bibfield  {title} {\bibinfo {title} {Quantum many-body phenomena in coupled
  cavity arrays},\ }\href@noop {} {\bibfield  {journal} {\bibinfo  {journal}
  {Laser Photonics Rev.}\ }\textbf {\bibinfo {volume} {2}},\ \bibinfo {pages}
  {527} (\bibinfo {year} {2008})}\BibitemShut {NoStop}%
\bibitem [{\citenamefont {Hohenadler}\ \emph {et~al.}(2012)\citenamefont
  {Hohenadler}, \citenamefont {Aichhorn}, \citenamefont {Pollet},\ and\
  \citenamefont {Schmidt}}]{PhysRevA.85.013810}%
  \BibitemOpen
  \bibfield  {author} {\bibinfo {author} {\bibfnamefont {M.}~\bibnamefont
  {Hohenadler}}, \bibinfo {author} {\bibfnamefont {M.}~\bibnamefont
  {Aichhorn}}, \bibinfo {author} {\bibfnamefont {L.}~\bibnamefont {Pollet}},\
  and\ \bibinfo {author} {\bibfnamefont {S.}~\bibnamefont {Schmidt}},\
  }\bibfield  {title} {\bibinfo {title} {Polariton {M}ott insulator with
  trapped ions or circuit {QED}},\ }\href
  {https://doi.org/10.1103/PhysRevA.85.013810} {\bibfield  {journal} {\bibinfo
  {journal} {Phys. Rev. A}\ }\textbf {\bibinfo {volume} {85}},\ \bibinfo
  {pages} {013810} (\bibinfo {year} {2012})}\BibitemShut {NoStop}%
\bibitem [{\citenamefont {Gessner}\ \emph {et~al.}(2015)\citenamefont
  {Gessner}, \citenamefont {Schlawin},\ and\ \citenamefont
  {Buchleitner}}]{Gessner2015}%
  \BibitemOpen
  \bibfield  {author} {\bibinfo {author} {\bibfnamefont {M.}~\bibnamefont
  {Gessner}}, \bibinfo {author} {\bibfnamefont {F.}~\bibnamefont {Schlawin}},\
  and\ \bibinfo {author} {\bibfnamefont {A.}~\bibnamefont {Buchleitner}},\
  }\bibfield  {title} {\bibinfo {title} {Probing polariton dynamics in trapped
  ions with phase-coherent two-dimensional spectroscopy},\ }\href@noop {}
  {\bibfield  {journal} {\bibinfo  {journal} {J. Chem. Phys.}\ }\textbf
  {\bibinfo {volume} {142}},\ \bibinfo {pages} {212439} (\bibinfo {year}
  {2015})}\BibitemShut {NoStop}%
\bibitem [{\citenamefont {Hartmann}(2016)}]{Hartmann2016}%
  \BibitemOpen
  \bibfield  {author} {\bibinfo {author} {\bibfnamefont {M.~J.}\ \bibnamefont
  {Hartmann}},\ }\bibfield  {title} {\bibinfo {title} {Quantum simulation with
  interacting photons},\ }\href@noop {} {\bibfield  {journal} {\bibinfo
  {journal} {J. Opt.}\ }\textbf {\bibinfo {volume} {18}},\ \bibinfo {pages}
  {104005} (\bibinfo {year} {2016})}\BibitemShut {NoStop}%
\bibitem [{\citenamefont {Chang}\ \emph {et~al.}(2018)\citenamefont {Chang},
  \citenamefont {Douglas}, \citenamefont {Gonzalez-Tudela}, \citenamefont
  {Hung},\ and\ \citenamefont {Kimble}}]{Chang2018}%
  \BibitemOpen
  \bibfield  {author} {\bibinfo {author} {\bibfnamefont {D.~E.}\ \bibnamefont
  {Chang}}, \bibinfo {author} {\bibfnamefont {J.~S.}\ \bibnamefont {Douglas}},
  \bibinfo {author} {\bibfnamefont {A.}~\bibnamefont {Gonzalez-Tudela}},
  \bibinfo {author} {\bibfnamefont {C.~L.}\ \bibnamefont {Hung}},\ and\
  \bibinfo {author} {\bibfnamefont {H.~J.}\ \bibnamefont {Kimble}},\ }\bibfield
   {title} {\bibinfo {title} {Colloquium: Quantum matter built from nanoscopic
  lattices of atoms and photons},\ }\href@noop {} {\bibfield  {journal}
  {\bibinfo  {journal} {Rev. Mod. Phys.}\ }\textbf {\bibinfo {volume} {90}},\
  \bibinfo {pages} {031002} (\bibinfo {year} {2018})}\BibitemShut {NoStop}%
\bibitem [{\citenamefont {Kato}\ \emph {et~al.}(2019)\citenamefont {Kato},
  \citenamefont {Nemet}, \citenamefont {Senga}, \citenamefont {Mizukami},
  \citenamefont {Huang}, \citenamefont {Parkins},\ and\ \citenamefont
  {Aoki}}]{Kato2019}%
  \BibitemOpen
  \bibfield  {author} {\bibinfo {author} {\bibfnamefont {S.}~\bibnamefont
  {Kato}}, \bibinfo {author} {\bibfnamefont {N.}~\bibnamefont {Nemet}},
  \bibinfo {author} {\bibfnamefont {K.}~\bibnamefont {Senga}}, \bibinfo
  {author} {\bibfnamefont {S.}~\bibnamefont {Mizukami}}, \bibinfo {author}
  {\bibfnamefont {X.~H.}\ \bibnamefont {Huang}}, \bibinfo {author}
  {\bibfnamefont {S.}~\bibnamefont {Parkins}},\ and\ \bibinfo {author}
  {\bibfnamefont {T.}~\bibnamefont {Aoki}},\ }\bibfield  {title} {\bibinfo
  {title} {Observation of dressed states of distant atoms with delocalized
  photons in coupled-cavities quantum electrodynamics},\ }\href@noop {}
  {\bibfield  {journal} {\bibinfo  {journal} {Nat. Commun.}\ }\textbf {\bibinfo
  {volume} {10}} (\bibinfo {year} {2019})}\BibitemShut {NoStop}%
\bibitem [{\citenamefont {Hoffman}\ \emph {et~al.}(2011)\citenamefont
  {Hoffman}, \citenamefont {Srinivasan}, \citenamefont {Schmidt}, \citenamefont
  {Spietz}, \citenamefont {Aumentado}, \citenamefont {Tureci},\ and\
  \citenamefont {Houck}}]{Hoffman2011}%
  \BibitemOpen
  \bibfield  {author} {\bibinfo {author} {\bibfnamefont {A.~J.}\ \bibnamefont
  {Hoffman}}, \bibinfo {author} {\bibfnamefont {S.~J.}\ \bibnamefont
  {Srinivasan}}, \bibinfo {author} {\bibfnamefont {S.}~\bibnamefont {Schmidt}},
  \bibinfo {author} {\bibfnamefont {L.}~\bibnamefont {Spietz}}, \bibinfo
  {author} {\bibfnamefont {J.}~\bibnamefont {Aumentado}}, \bibinfo {author}
  {\bibfnamefont {H.~E.}\ \bibnamefont {Tureci}},\ and\ \bibinfo {author}
  {\bibfnamefont {A.~A.}\ \bibnamefont {Houck}},\ }\bibfield  {title} {\bibinfo
  {title} {Dispersive photon blockade in a superconducting circuit},\
  }\href@noop {} {\bibfield  {journal} {\bibinfo  {journal} {Phys. Rev. Lett.}\
  }\textbf {\bibinfo {volume} {107}},\ \bibinfo {pages} {053602} (\bibinfo
  {year} {2011})}\BibitemShut {NoStop}%
\bibitem [{\citenamefont {Houck}\ \emph {et~al.}(2012)\citenamefont {Houck},
  \citenamefont {Tureci},\ and\ \citenamefont {Koch}}]{Houck2012}%
  \BibitemOpen
  \bibfield  {author} {\bibinfo {author} {\bibfnamefont {A.~A.}\ \bibnamefont
  {Houck}}, \bibinfo {author} {\bibfnamefont {H.~E.}\ \bibnamefont {Tureci}},\
  and\ \bibinfo {author} {\bibfnamefont {J.}~\bibnamefont {Koch}},\ }\bibfield
  {title} {\bibinfo {title} {On-chip quantum simulation with superconducting
  circuits},\ }\href@noop {} {\bibfield  {journal} {\bibinfo  {journal} {Nat.
  Phys.}\ }\textbf {\bibinfo {volume} {8}},\ \bibinfo {pages} {292} (\bibinfo
  {year} {2012})}\BibitemShut {NoStop}%
\bibitem [{\citenamefont {Schmidt}\ and\ \citenamefont
  {Koch}(2013)}]{Schmidt2013}%
  \BibitemOpen
  \bibfield  {author} {\bibinfo {author} {\bibfnamefont {S.}~\bibnamefont
  {Schmidt}}\ and\ \bibinfo {author} {\bibfnamefont {J.}~\bibnamefont {Koch}},\
  }\bibfield  {title} {\bibinfo {title} {Circuit qed lattices: Towards quantum
  simulation with superconducting circuits},\ }\href@noop {} {\bibfield
  {journal} {\bibinfo  {journal} {Ann. Phys.}\ }\textbf {\bibinfo {volume}
  {525}},\ \bibinfo {pages} {395} (\bibinfo {year} {2013})}\BibitemShut
  {NoStop}%
\bibitem [{\citenamefont {Raftery}\ \emph {et~al.}(2014)\citenamefont
  {Raftery}, \citenamefont {Sadri}, \citenamefont {Schmidt}, \citenamefont
  {Tureci},\ and\ \citenamefont {Houck}}]{Raftery2014}%
  \BibitemOpen
  \bibfield  {author} {\bibinfo {author} {\bibfnamefont {J.}~\bibnamefont
  {Raftery}}, \bibinfo {author} {\bibfnamefont {D.}~\bibnamefont {Sadri}},
  \bibinfo {author} {\bibfnamefont {S.}~\bibnamefont {Schmidt}}, \bibinfo
  {author} {\bibfnamefont {H.~E.}\ \bibnamefont {Tureci}},\ and\ \bibinfo
  {author} {\bibfnamefont {A.~A.}\ \bibnamefont {Houck}},\ }\bibfield  {title}
  {\bibinfo {title} {Observation of a dissipation-induced classical to quantum
  transition},\ }\href@noop {} {\bibfield  {journal} {\bibinfo  {journal}
  {Phys. Rev. X}\ }\textbf {\bibinfo {volume} {4}},\ \bibinfo {pages} {031043}
  (\bibinfo {year} {2014})}\BibitemShut {NoStop}%
\bibitem [{\citenamefont {Fitzpatrick}\ \emph {et~al.}(2017)\citenamefont
  {Fitzpatrick}, \citenamefont {Sundaresan}, \citenamefont {Li}, \citenamefont
  {Koch},\ and\ \citenamefont {Houck}}]{Fitzpatrick2017}%
  \BibitemOpen
  \bibfield  {author} {\bibinfo {author} {\bibfnamefont {M.}~\bibnamefont
  {Fitzpatrick}}, \bibinfo {author} {\bibfnamefont {N.~M.}\ \bibnamefont
  {Sundaresan}}, \bibinfo {author} {\bibfnamefont {A.~C.~Y.}\ \bibnamefont
  {Li}}, \bibinfo {author} {\bibfnamefont {J.}~\bibnamefont {Koch}},\ and\
  \bibinfo {author} {\bibfnamefont {A.~A.}\ \bibnamefont {Houck}},\ }\bibfield
  {title} {\bibinfo {title} {Observation of a dissipative phase transition in a
  one-dimensional circuit qed lattice},\ }\href@noop {} {\bibfield  {journal}
  {\bibinfo  {journal} {Phys. Rev. X}\ }\textbf {\bibinfo {volume} {7}},\
  \bibinfo {pages} {011016} (\bibinfo {year} {2017})}\BibitemShut {NoStop}%
\bibitem [{\citenamefont {Ivanov}\ \emph {et~al.}(2009)\citenamefont {Ivanov},
  \citenamefont {Ivanov}, \citenamefont {Vitanov}, \citenamefont {Mering},
  \citenamefont {Fleischhauer},\ and\ \citenamefont {Singer}}]{Ivanov2009}%
  \BibitemOpen
  \bibfield  {author} {\bibinfo {author} {\bibfnamefont {P.~A.}\ \bibnamefont
  {Ivanov}}, \bibinfo {author} {\bibfnamefont {S.~S.}\ \bibnamefont {Ivanov}},
  \bibinfo {author} {\bibfnamefont {N.~V.}\ \bibnamefont {Vitanov}}, \bibinfo
  {author} {\bibfnamefont {A.}~\bibnamefont {Mering}}, \bibinfo {author}
  {\bibfnamefont {M.}~\bibnamefont {Fleischhauer}},\ and\ \bibinfo {author}
  {\bibfnamefont {K.}~\bibnamefont {Singer}},\ }\bibfield  {title} {\bibinfo
  {title} {Simulation of a quantum phase transition of polaritons with trapped
  ions},\ }\href@noop {} {\bibfield  {journal} {\bibinfo  {journal} {Phys. Rev.
  A}\ }\textbf {\bibinfo {volume} {80}},\ \bibinfo {pages} {060301(R)}
  (\bibinfo {year} {2009})}\BibitemShut {NoStop}%
\bibitem [{\citenamefont {Toyoda}\ \emph {et~al.}(2013)\citenamefont {Toyoda},
  \citenamefont {Matsuno}, \citenamefont {Noguchi}, \citenamefont {Haze},\ and\
  \citenamefont {Urabe}}]{Toyo2013}%
  \BibitemOpen
  \bibfield  {author} {\bibinfo {author} {\bibfnamefont {K.}~\bibnamefont
  {Toyoda}}, \bibinfo {author} {\bibfnamefont {Y.}~\bibnamefont {Matsuno}},
  \bibinfo {author} {\bibfnamefont {A.}~\bibnamefont {Noguchi}}, \bibinfo
  {author} {\bibfnamefont {S.}~\bibnamefont {Haze}},\ and\ \bibinfo {author}
  {\bibfnamefont {S.}~\bibnamefont {Urabe}},\ }\bibfield  {title} {\bibinfo
  {title} {Experimental realization of a quantum phase transition of
  polaritonic excitation},\ }\href@noop {} {\bibfield  {journal} {\bibinfo
  {journal} {Phys. Rev. Lett.}\ }\textbf {\bibinfo {volume} {111}},\ \bibinfo
  {pages} {160501} (\bibinfo {year} {2013})}\BibitemShut {NoStop}%
\bibitem [{\citenamefont {Debnath}\ \emph {et~al.}(2018)\citenamefont
  {Debnath}, \citenamefont {Linke}, \citenamefont {Wang}, \citenamefont
  {Figgatt}, \citenamefont {Landsman}, \citenamefont {Duan},\ and\
  \citenamefont {Monroe}}]{PhysRevLett.120.073001}%
  \BibitemOpen
  \bibfield  {author} {\bibinfo {author} {\bibfnamefont {S.}~\bibnamefont
  {Debnath}}, \bibinfo {author} {\bibfnamefont {N.~M.}\ \bibnamefont {Linke}},
  \bibinfo {author} {\bibfnamefont {S.-T.}\ \bibnamefont {Wang}}, \bibinfo
  {author} {\bibfnamefont {C.}~\bibnamefont {Figgatt}}, \bibinfo {author}
  {\bibfnamefont {K.~A.}\ \bibnamefont {Landsman}}, \bibinfo {author}
  {\bibfnamefont {L.-M.}\ \bibnamefont {Duan}},\ and\ \bibinfo {author}
  {\bibfnamefont {C.}~\bibnamefont {Monroe}},\ }\bibfield  {title} {\bibinfo
  {title} {Observation of hopping and blockade of bosons in a trapped ion spin
  chain},\ }\href {https://doi.org/10.1103/PhysRevLett.120.073001} {\bibfield
  {journal} {\bibinfo  {journal} {Phys. Rev. Lett.}\ }\textbf {\bibinfo
  {volume} {120}},\ \bibinfo {pages} {073001} (\bibinfo {year}
  {2018})}\BibitemShut {NoStop}%
\bibitem [{\citenamefont {Ohira}\ \emph
  {et~al.}(2021{\natexlab{a}})\citenamefont {Ohira}, \citenamefont {Kume},
  \citenamefont {Takayama}, \citenamefont {Muralidharan}, \citenamefont
  {Takahashi},\ and\ \citenamefont {Toyoda}}]{PhysRevA.103.012612}%
  \BibitemOpen
  \bibfield  {author} {\bibinfo {author} {\bibfnamefont {R.}~\bibnamefont
  {Ohira}}, \bibinfo {author} {\bibfnamefont {S.}~\bibnamefont {Kume}},
  \bibinfo {author} {\bibfnamefont {K.}~\bibnamefont {Takayama}}, \bibinfo
  {author} {\bibfnamefont {S.}~\bibnamefont {Muralidharan}}, \bibinfo {author}
  {\bibfnamefont {H.}~\bibnamefont {Takahashi}},\ and\ \bibinfo {author}
  {\bibfnamefont {K.}~\bibnamefont {Toyoda}},\ }\bibfield  {title} {\bibinfo
  {title} {Blockade of phonon hopping in trapped ions in the presence of
  multiple local phonons},\ }\href
  {https://doi.org/10.1103/PhysRevA.103.012612} {\bibfield  {journal} {\bibinfo
   {journal} {Phys. Rev. A}\ }\textbf {\bibinfo {volume} {103}},\ \bibinfo
  {pages} {012612} (\bibinfo {year} {2021}{\natexlab{a}})}\BibitemShut
  {NoStop}%
\bibitem [{\citenamefont {Ohira}\ \emph
  {et~al.}(2021{\natexlab{b}})\citenamefont {Ohira}, \citenamefont {Kume},
  \citenamefont {Takahashi},\ and\ \citenamefont
  {Toyoda}}]{ohira2021polariton}%
  \BibitemOpen
  \bibfield  {author} {\bibinfo {author} {\bibfnamefont {R.}~\bibnamefont
  {Ohira}}, \bibinfo {author} {\bibfnamefont {S.}~\bibnamefont {Kume}},
  \bibinfo {author} {\bibfnamefont {H.}~\bibnamefont {Takahashi}},\ and\
  \bibinfo {author} {\bibfnamefont {K.}~\bibnamefont {Toyoda}},\ }\bibfield
  {title} {\bibinfo {title} {Polariton blockade in the
  {J}aynes-{C}ummings-{H}ubbard model with trapped ions},\ }\href@noop {}
  {\bibfield  {journal} {\bibinfo  {journal} {Quantum Sci. Technol.}\ }\textbf
  {\bibinfo {volume} {6}},\ \bibinfo {pages} {024015} (\bibinfo {year}
  {2021}{\natexlab{b}})}\BibitemShut {NoStop}%
\bibitem [{\citenamefont {Porras}\ and\ \citenamefont
  {Cirac}(2004)}]{Porras2004}%
  \BibitemOpen
  \bibfield  {author} {\bibinfo {author} {\bibfnamefont {D.}~\bibnamefont
  {Porras}}\ and\ \bibinfo {author} {\bibfnamefont {J.~I.}\ \bibnamefont
  {Cirac}},\ }\bibfield  {title} {\bibinfo {title} {Bose-{E}instein
  condensation and strong-correlation behavior of phonons in ion traps},\
  }\href@noop {} {\bibfield  {journal} {\bibinfo  {journal} {Phys. Rev. Lett.}\
  }\textbf {\bibinfo {volume} {93}},\ \bibinfo {pages} {263602} (\bibinfo
  {year} {2004})}\BibitemShut {NoStop}%
\bibitem [{\citenamefont {Brown}\ \emph {et~al.}(2011)\citenamefont {Brown},
  \citenamefont {Ospelkaus}, \citenamefont {Colombe}, \citenamefont {Wilson},
  \citenamefont {Leibfried},\ and\ \citenamefont {Wineland}}]{Brown2011}%
  \BibitemOpen
  \bibfield  {author} {\bibinfo {author} {\bibfnamefont {K.~R.}\ \bibnamefont
  {Brown}}, \bibinfo {author} {\bibfnamefont {C.}~\bibnamefont {Ospelkaus}},
  \bibinfo {author} {\bibfnamefont {Y.}~\bibnamefont {Colombe}}, \bibinfo
  {author} {\bibfnamefont {A.~C.}\ \bibnamefont {Wilson}}, \bibinfo {author}
  {\bibfnamefont {D.}~\bibnamefont {Leibfried}},\ and\ \bibinfo {author}
  {\bibfnamefont {D.~J.}\ \bibnamefont {Wineland}},\ }\bibfield  {title}
  {\bibinfo {title} {Coupled quantized mechanical oscillators},\ }\href@noop {}
  {\bibfield  {journal} {\bibinfo  {journal} {Nature (London)}\ }\textbf
  {\bibinfo {volume} {471}},\ \bibinfo {pages} {196} (\bibinfo {year}
  {2011})}\BibitemShut {NoStop}%
\bibitem [{\citenamefont {Harlander}\ \emph {et~al.}(2011)\citenamefont
  {Harlander}, \citenamefont {Lechner}, \citenamefont {Brownnutt},
  \citenamefont {Blatt},\ and\ \citenamefont {H\"ansel}}]{Harlander2011}%
  \BibitemOpen
  \bibfield  {author} {\bibinfo {author} {\bibfnamefont {M.}~\bibnamefont
  {Harlander}}, \bibinfo {author} {\bibfnamefont {R.}~\bibnamefont {Lechner}},
  \bibinfo {author} {\bibfnamefont {M.}~\bibnamefont {Brownnutt}}, \bibinfo
  {author} {\bibfnamefont {R.}~\bibnamefont {Blatt}},\ and\ \bibinfo {author}
  {\bibfnamefont {W.}~\bibnamefont {H\"ansel}},\ }\bibfield  {title} {\bibinfo
  {title} {Trapped-ion antennae for the transmission of quantum information},\
  }\href@noop {} {\bibfield  {journal} {\bibinfo  {journal} {Nature (London)}\
  }\textbf {\bibinfo {volume} {471}},\ \bibinfo {pages} {200} (\bibinfo {year}
  {2011})}\BibitemShut {NoStop}%
\bibitem [{\citenamefont {Haze}\ \emph {et~al.}(2012)\citenamefont {Haze},
  \citenamefont {Tateishi}, \citenamefont {Noguchi}, \citenamefont {Toyoda},\
  and\ \citenamefont {Urabe}}]{Haze2012}%
  \BibitemOpen
  \bibfield  {author} {\bibinfo {author} {\bibfnamefont {S.}~\bibnamefont
  {Haze}}, \bibinfo {author} {\bibfnamefont {Y.}~\bibnamefont {Tateishi}},
  \bibinfo {author} {\bibfnamefont {A.}~\bibnamefont {Noguchi}}, \bibinfo
  {author} {\bibfnamefont {K.}~\bibnamefont {Toyoda}},\ and\ \bibinfo {author}
  {\bibfnamefont {S.}~\bibnamefont {Urabe}},\ }\bibfield  {title} {\bibinfo
  {title} {Observation of phonon hopping in radial vibrational modes of trapped
  ions},\ }\href@noop {} {\bibfield  {journal} {\bibinfo  {journal} {Phys. Rev.
  A}\ }\textbf {\bibinfo {volume} {85}},\ \bibinfo {pages} {031401(R)}
  (\bibinfo {year} {2012})}\BibitemShut {NoStop}%
\bibitem [{\citenamefont {Wilson}\ \emph {et~al.}(2014)\citenamefont {Wilson},
  \citenamefont {Colombe}, \citenamefont {Brown}, \citenamefont {Knill},
  \citenamefont {Leibfried},\ and\ \citenamefont
  {Wineland}}]{wilson2014tunable}%
  \BibitemOpen
  \bibfield  {author} {\bibinfo {author} {\bibfnamefont {A.~C.}\ \bibnamefont
  {Wilson}}, \bibinfo {author} {\bibfnamefont {Y.}~\bibnamefont {Colombe}},
  \bibinfo {author} {\bibfnamefont {K.~R.}\ \bibnamefont {Brown}}, \bibinfo
  {author} {\bibfnamefont {E.}~\bibnamefont {Knill}}, \bibinfo {author}
  {\bibfnamefont {D.}~\bibnamefont {Leibfried}},\ and\ \bibinfo {author}
  {\bibfnamefont {D.~J.}\ \bibnamefont {Wineland}},\ }\bibfield  {title}
  {\bibinfo {title} {Tunable spin--spin interactions and entanglement of ions
  in separate potential wells},\ }\href@noop {} {\bibfield  {journal} {\bibinfo
   {journal} {Nature}\ }\textbf {\bibinfo {volume} {512}},\ \bibinfo {pages}
  {57} (\bibinfo {year} {2014})}\BibitemShut {NoStop}%
\bibitem [{\citenamefont {Toyoda}\ \emph {et~al.}(2015)\citenamefont {Toyoda},
  \citenamefont {Hiji}, \citenamefont {Noguchi},\ and\ \citenamefont
  {Urabe}}]{toyoda2015hong}%
  \BibitemOpen
  \bibfield  {author} {\bibinfo {author} {\bibfnamefont {K.}~\bibnamefont
  {Toyoda}}, \bibinfo {author} {\bibfnamefont {R.}~\bibnamefont {Hiji}},
  \bibinfo {author} {\bibfnamefont {A.}~\bibnamefont {Noguchi}},\ and\ \bibinfo
  {author} {\bibfnamefont {S.}~\bibnamefont {Urabe}},\ }\bibfield  {title}
  {\bibinfo {title} {Hong--{O}u--{M}andel interference of two phonons in
  trapped ions},\ }\href@noop {} {\bibfield  {journal} {\bibinfo  {journal}
  {Nature}\ }\textbf {\bibinfo {volume} {527}},\ \bibinfo {pages} {74}
  (\bibinfo {year} {2015})}\BibitemShut {NoStop}%
\bibitem [{\citenamefont {Tamura}\ \emph {et~al.}(2020)\citenamefont {Tamura},
  \citenamefont {Mukaiyama},\ and\ \citenamefont
  {Toyoda}}]{PhysRevLett.124.200501}%
  \BibitemOpen
  \bibfield  {author} {\bibinfo {author} {\bibfnamefont {M.}~\bibnamefont
  {Tamura}}, \bibinfo {author} {\bibfnamefont {T.}~\bibnamefont {Mukaiyama}},\
  and\ \bibinfo {author} {\bibfnamefont {K.}~\bibnamefont {Toyoda}},\
  }\bibfield  {title} {\bibinfo {title} {Quantum walks of a phonon in trapped
  ions},\ }\href {https://doi.org/10.1103/PhysRevLett.124.200501} {\bibfield
  {journal} {\bibinfo  {journal} {Phys. Rev. Lett.}\ }\textbf {\bibinfo
  {volume} {124}},\ \bibinfo {pages} {200501} (\bibinfo {year}
  {2020})}\BibitemShut {NoStop}%
\bibitem [{\citenamefont {Ohira}\ \emph {et~al.}(2019)\citenamefont {Ohira},
  \citenamefont {Mukaiyama},\ and\ \citenamefont
  {Toyoda}}]{PhysRevA.100.060301}%
  \BibitemOpen
  \bibfield  {author} {\bibinfo {author} {\bibfnamefont {R.}~\bibnamefont
  {Ohira}}, \bibinfo {author} {\bibfnamefont {T.}~\bibnamefont {Mukaiyama}},\
  and\ \bibinfo {author} {\bibfnamefont {K.}~\bibnamefont {Toyoda}},\
  }\bibfield  {title} {\bibinfo {title} {Phonon-number-resolving detection of
  multiple local phonon modes in trapped ions},\ }\href
  {https://doi.org/10.1103/PhysRevA.100.060301} {\bibfield  {journal} {\bibinfo
   {journal} {Phys. Rev. A}\ }\textbf {\bibinfo {volume} {100}},\ \bibinfo
  {pages} {060301} (\bibinfo {year} {2019})}\BibitemShut {NoStop}%
\bibitem [{\citenamefont {Wang}\ \emph {et~al.}(2010)\citenamefont {Wang},
  \citenamefont {Labaziewicz}, \citenamefont {Ge}, \citenamefont {Shewmon},\
  and\ \citenamefont {Chuang}}]{Wang2010}%
  \BibitemOpen
  \bibfield  {author} {\bibinfo {author} {\bibfnamefont {S.~X.}\ \bibnamefont
  {Wang}}, \bibinfo {author} {\bibfnamefont {J.}~\bibnamefont {Labaziewicz}},
  \bibinfo {author} {\bibfnamefont {Y.}~\bibnamefont {Ge}}, \bibinfo {author}
  {\bibfnamefont {R.}~\bibnamefont {Shewmon}},\ and\ \bibinfo {author}
  {\bibfnamefont {I.~L.}\ \bibnamefont {Chuang}},\ }\bibfield  {title}
  {\bibinfo {title} {Demonstration of a quantum logic gate in a cryogenic
  surface-electrode ion trap},\ }\href@noop {} {\bibfield  {journal} {\bibinfo
  {journal} {Phys. Rev. A}\ }\textbf {\bibinfo {volume} {81}},\ \bibinfo
  {pages} {062332} (\bibinfo {year} {2010})}\BibitemShut {NoStop}%
\bibitem [{\citenamefont {Vitanov}\ \emph {et~al.}(2001)\citenamefont
  {Vitanov}, \citenamefont {halfmann}, \citenamefont {Shore},\ and\
  \citenamefont {Bergmann}}]{Nikolay2001}%
  \BibitemOpen
  \bibfield  {author} {\bibinfo {author} {\bibfnamefont {N.~V.}\ \bibnamefont
  {Vitanov}}, \bibinfo {author} {\bibfnamefont {T.}~\bibnamefont {halfmann}},
  \bibinfo {author} {\bibfnamefont {B.~W.}\ \bibnamefont {Shore}},\ and\
  \bibinfo {author} {\bibfnamefont {K.}~\bibnamefont {Bergmann}},\ }\bibfield
  {title} {\bibinfo {title} {Laser-induced population transfer by adiabatic
  passage technique},\ }\href@noop {} {\bibfield  {journal} {\bibinfo
  {journal} {Annu. Rev. Phys. Chem.}\ }\textbf {\bibinfo {volume} {52}},\
  \bibinfo {pages} {763} (\bibinfo {year} {2001})}\BibitemShut {NoStop}%
\bibitem [{\citenamefont {Irish}\ \emph {et~al.}(2008)\citenamefont {Irish},
  \citenamefont {Ogden},\ and\ \citenamefont {Kim}}]{Irish2008}%
  \BibitemOpen
  \bibfield  {author} {\bibinfo {author} {\bibfnamefont {E.~K.}\ \bibnamefont
  {Irish}}, \bibinfo {author} {\bibfnamefont {C.~D.}\ \bibnamefont {Ogden}},\
  and\ \bibinfo {author} {\bibfnamefont {M.~S.}\ \bibnamefont {Kim}},\
  }\bibfield  {title} {\bibinfo {title} {Polaritonic characteristics of
  insulator and superfluid states in a coupled-cavity array},\ }\href@noop {}
  {\bibfield  {journal} {\bibinfo  {journal} {Phys. Rev. A}\ }\textbf {\bibinfo
  {volume} {77}},\ \bibinfo {pages} {033801} (\bibinfo {year}
  {2008})}\BibitemShut {NoStop}%
\bibitem [{\citenamefont {Jaksch}\ \emph {et~al.}(1998)\citenamefont {Jaksch},
  \citenamefont {Bruder}, \citenamefont {Cirac}, \citenamefont {Gardiner},\
  and\ \citenamefont {Zoller}}]{Jaksch1998}%
  \BibitemOpen
  \bibfield  {author} {\bibinfo {author} {\bibfnamefont {D.}~\bibnamefont
  {Jaksch}}, \bibinfo {author} {\bibfnamefont {C.}~\bibnamefont {Bruder}},
  \bibinfo {author} {\bibfnamefont {J.~I.}\ \bibnamefont {Cirac}}, \bibinfo
  {author} {\bibfnamefont {C.~W.}\ \bibnamefont {Gardiner}},\ and\ \bibinfo
  {author} {\bibfnamefont {P.}~\bibnamefont {Zoller}},\ }\bibfield  {title}
  {\bibinfo {title} {Cold bosonic atoms in optical lattices},\ }\href {<Go to
  ISI>://WOS:000076369400015 http://prl.aps.org/pdf/PRL/v81/i15/p3108_1}
  {\bibfield  {journal} {\bibinfo  {journal} {Phys. Rev. Lett.}\ }\textbf
  {\bibinfo {volume} {81}},\ \bibinfo {pages} {3108} (\bibinfo {year}
  {1998})}\BibitemShut {NoStop}%
\bibitem [{\citenamefont {Greiner}\ \emph {et~al.}(2002)\citenamefont
  {Greiner}, \citenamefont {Mandel}, \citenamefont {Esslinger}, \citenamefont
  {Hansch},\ and\ \citenamefont {Bloch}}]{greiner2002}%
  \BibitemOpen
  \bibfield  {author} {\bibinfo {author} {\bibfnamefont {M.}~\bibnamefont
  {Greiner}}, \bibinfo {author} {\bibfnamefont {O.}~\bibnamefont {Mandel}},
  \bibinfo {author} {\bibfnamefont {T.}~\bibnamefont {Esslinger}}, \bibinfo
  {author} {\bibfnamefont {T.~W.}\ \bibnamefont {Hansch}},\ and\ \bibinfo
  {author} {\bibfnamefont {I.}~\bibnamefont {Bloch}},\ }\bibfield  {title}
  {\bibinfo {title} {Quantum phase transition from a superfluid to a {M}ott
  insulator in a gas of ultracold atoms},\ }\href@noop {} {\bibfield  {journal}
  {\bibinfo  {journal} {Nature (London)}\ }\textbf {\bibinfo {volume} {415}},\
  \bibinfo {pages} {937} (\bibinfo {year} {2002})}\BibitemShut {NoStop}%
\bibitem [{\citenamefont {Deng}\ \emph {et~al.}(2008)\citenamefont {Deng},
  \citenamefont {Porras},\ and\ \citenamefont {Cirac}}]{Deng2008}%
  \BibitemOpen
  \bibfield  {author} {\bibinfo {author} {\bibfnamefont {X.~L.}\ \bibnamefont
  {Deng}}, \bibinfo {author} {\bibfnamefont {D.}~\bibnamefont {Porras}},\ and\
  \bibinfo {author} {\bibfnamefont {J.~I.}\ \bibnamefont {Cirac}},\ }\bibfield
  {title} {\bibinfo {title} {Quantum phases of interacting phonons in ion
  traps},\ }\href@noop {} {\bibfield  {journal} {\bibinfo  {journal} {Physical
  Review A}\ }\textbf {\bibinfo {volume} {77}},\ \bibinfo {pages} {033403}
  (\bibinfo {year} {2008})}\BibitemShut {NoStop}%
\bibitem [{\citenamefont {Wineland}\ \emph {et~al.}(1998)\citenamefont
  {Wineland}, \citenamefont {Monroe}, \citenamefont {Itano}, \citenamefont
  {Leibfried}, \citenamefont {King},\ and\ \citenamefont
  {Meekhof}}]{Wineland1998}%
  \BibitemOpen
  \bibfield  {author} {\bibinfo {author} {\bibfnamefont {D.~J.}\ \bibnamefont
  {Wineland}}, \bibinfo {author} {\bibfnamefont {C.}~\bibnamefont {Monroe}},
  \bibinfo {author} {\bibfnamefont {W.~M.}\ \bibnamefont {Itano}}, \bibinfo
  {author} {\bibfnamefont {D.}~\bibnamefont {Leibfried}}, \bibinfo {author}
  {\bibfnamefont {B.~E.}\ \bibnamefont {King}},\ and\ \bibinfo {author}
  {\bibfnamefont {D.~M.}\ \bibnamefont {Meekhof}},\ }\bibfield  {title}
  {\bibinfo {title} {Experimental issues in coherent quantum-state manipulation
  of trapped atomic ions},\ }\href@noop {} {\bibfield  {journal} {\bibinfo
  {journal} {J. Res. Natl. Inst. Stand. Technol.}\ }\textbf {\bibinfo {volume}
  {103}},\ \bibinfo {pages} {259} (\bibinfo {year} {1998})}\BibitemShut
  {NoStop}%
\bibitem [{\citenamefont {Šašura}\ and\ \citenamefont
  {Bužek}(2002)}]{Sasura2001}%
  \BibitemOpen
  \bibfield  {author} {\bibinfo {author} {\bibfnamefont {M.}~\bibnamefont
  {Šašura}}\ and\ \bibinfo {author} {\bibfnamefont {V.}~\bibnamefont
  {Bužek}},\ }\bibfield  {title} {\bibinfo {title} {Cold trapped ions as
  quantum information processors},\ }\href@noop {} {\bibfield  {journal}
  {\bibinfo  {journal} {J. Mod. Opt.}\ }\textbf {\bibinfo {volume} {49}},\
  \bibinfo {pages} {1593} (\bibinfo {year} {2002})}\BibitemShut {NoStop}%
\bibitem [{\citenamefont {Johnson}\ \emph {et~al.}(2016)\citenamefont
  {Johnson}, \citenamefont {Wong-Campos}, \citenamefont {Restelli},
  \citenamefont {Landsman}, \citenamefont {Neyenhuis}, \citenamefont
  {Mizrahi},\ and\ \citenamefont {Monroe}}]{Johnson2016}%
  \BibitemOpen
  \bibfield  {author} {\bibinfo {author} {\bibfnamefont {K.~G.}\ \bibnamefont
  {Johnson}}, \bibinfo {author} {\bibfnamefont {J.~D.}\ \bibnamefont
  {Wong-Campos}}, \bibinfo {author} {\bibfnamefont {A.}~\bibnamefont
  {Restelli}}, \bibinfo {author} {\bibfnamefont {K.~A.}\ \bibnamefont
  {Landsman}}, \bibinfo {author} {\bibfnamefont {B.}~\bibnamefont {Neyenhuis}},
  \bibinfo {author} {\bibfnamefont {J.}~\bibnamefont {Mizrahi}},\ and\ \bibinfo
  {author} {\bibfnamefont {C.}~\bibnamefont {Monroe}},\ }\bibfield  {title}
  {\bibinfo {title} {Active stabilization of ion trap radiofrequency
  potentials},\ }\href {https://doi.org/Artn 053110 10.1063/1.4948734}
  {\bibfield  {journal} {\bibinfo  {journal} {Rev. Sci. Instrum.}\ }\textbf
  {\bibinfo {volume} {87}},\ \bibinfo {pages} {053110} (\bibinfo {year}
  {2016})}\BibitemShut {NoStop}%
\bibitem [{\citenamefont {Manzano}(2020)}]{Manzano2020}%
  \BibitemOpen
  \bibfield  {author} {\bibinfo {author} {\bibfnamefont {D.}~\bibnamefont
  {Manzano}},\ }\bibfield  {title} {\bibinfo {title} {A short introduction to
  the lindblad master equation},\ }\href@noop {} {\bibfield  {journal}
  {\bibinfo  {journal} {AIP Adv.}\ }\textbf {\bibinfo {volume} {10}},\ \bibinfo
  {pages} {025106} (\bibinfo {year} {2020})}\BibitemShut {NoStop}%
\bibitem [{Note1()}]{Note1}%
  \BibitemOpen
  \bibinfo {note} {The cause of the reduction by 5 \% in the population of
  $\mathinner {|{e}\rangle }$ can be explained as being due to the incomplete
  preparation of the atomic MI state by the carrier $\pi $ pulse. We use the
  individual illumination of each ion with a dedicated beam at 729 nm. Each
  beam is passed through a dedicated polarization-maintaining optical fiber to
  assure path stability. However, this fiber may produce a drift in the optical
  phases. Fluctuations in interference between the beams caused by this drift
  of optical phases, as well as those in the pointing of the beams, lead to
  relatively large intensity fluctuations at the positions of the ions and
  affect the fidelity of rotations even for the carrier
  transition.}\BibitemShut {Stop}%
\bibitem [{\citenamefont {Meekhof}\ \emph {et~al.}(1996)\citenamefont
  {Meekhof}, \citenamefont {Monroe}, \citenamefont {King}, \citenamefont
  {Itano},\ and\ \citenamefont {Wineland}}]{Meekhof1996}%
  \BibitemOpen
  \bibfield  {author} {\bibinfo {author} {\bibfnamefont {D.}~\bibnamefont
  {Meekhof}}, \bibinfo {author} {\bibfnamefont {C.}~\bibnamefont {Monroe}},
  \bibinfo {author} {\bibfnamefont {B.}~\bibnamefont {King}}, \bibinfo {author}
  {\bibfnamefont {W.}~\bibnamefont {Itano}},\ and\ \bibinfo {author}
  {\bibfnamefont {D.}~\bibnamefont {Wineland}},\ }\bibfield  {title} {\bibinfo
  {title} {Generation of nonclassical motional states of a trapped atom},\
  }\href@noop {} {\bibfield  {journal} {\bibinfo  {journal} {Phys. Rev. Lett.}\
  }\textbf {\bibinfo {volume} {76}},\ \bibinfo {pages} {1792} (\bibinfo {year}
  {1996})}\BibitemShut {NoStop}%
\bibitem [{Note2()}]{Note2}%
  \BibitemOpen
  \bibinfo {note} {In principle, the previously quoted transverse-relaxation
  rate $\gamma _\protect \mathrm {T}$ in the Lindblad master equation and this
  $\gamma _\protect \mathrm {R}$ should have a certain relation. If the latter
  is assumed to be determined purely from the former, $\gamma _\protect \mathrm
  {R}\sim 0.25\times \gamma _\protect \mathrm {T}$ is expected, which is
  confirmed by numerical simulations. In that case, $\gamma _\protect \mathrm
  {R}\sim 0.05$ kHz is expected if $\gamma _\protect \mathrm {T}\sim 2\pi
  \times 0.19$ kHz is satisfied. On the other hand, the actual value of $\gamma
  _\protect \mathrm {R}$ determined from the chi-square minimization, as
  explained in Sec. \ref {sec:VIIa}, is $2\pi \times 1.1$ kHz, which is more
  than an order of magnitude larger than the above-mentioned value. This
  discrepancy may be explained by the concrete nature of the fluctuations. If
  an adiabatic-transfer process like that in Fig.~\ref {rap}(a) is used and the
  typical timescale for the fluctuations is much longer than the pulse
  duration, the adiabatic transfer process is not affected significantly by the
  fluctuations thanks to their robustness against parameter variations. This
  may be why the assumed relaxation rate that matches the experimental results
  is much smaller than that for the numerical simulation results shown in
  Fig.~\ref {rap}(a).}\BibitemShut {Stop}%
\bibitem [{Note3()}]{Note3}%
  \BibitemOpen
  \bibinfo {note} {The factors that may make this assumption invalid are the
  residual phonon number after sideband cooling (estimated to be $\sim 0.2$ in
  the present case), heating during adiabatic transfer ($<\sim 5\times 10^{-3}$
  quanta), and off-resonance excitation of the carrier transition by the
  red-sideband optical pulse for the adiabatic transfer ($<0.03$). Here, we
  ignore their effects.}\BibitemShut {Stop}%
\bibitem [{Note4()}]{Note4}%
  \BibitemOpen
  \bibinfo {note} {Here, for simplicity, we assume that the errors are
  symmetric for the positive and negative directions, and nonlinear dependences
  of the deviations around the estimated values are ignored. This could lead to
  the appearance of unphysical values for the bounds of the confidence
  intervals: the negative values in Fig.~\ref {sbvar}(a) and (b) at 96 and 192
  $\mu $s are considered to be due to this.}\BibitemShut {Stop}%
\bibitem [{\citenamefont {Islam}\ \emph {et~al.}(2011)\citenamefont {Islam},
  \citenamefont {Edwards}, \citenamefont {Kim}, \citenamefont {Korenblit},
  \citenamefont {Noh}, \citenamefont {Carmichael}, \citenamefont {Lin},
  \citenamefont {Duan}, \citenamefont {Joseph~Wang}, \citenamefont
  {Freericks},\ and\ \citenamefont {Monroe}}]{Islam2011}%
  \BibitemOpen
  \bibfield  {author} {\bibinfo {author} {\bibfnamefont {R.}~\bibnamefont
  {Islam}}, \bibinfo {author} {\bibfnamefont {E.~E.}\ \bibnamefont {Edwards}},
  \bibinfo {author} {\bibfnamefont {K.}~\bibnamefont {Kim}}, \bibinfo {author}
  {\bibfnamefont {S.}~\bibnamefont {Korenblit}}, \bibinfo {author}
  {\bibfnamefont {C.}~\bibnamefont {Noh}}, \bibinfo {author} {\bibfnamefont
  {H.}~\bibnamefont {Carmichael}}, \bibinfo {author} {\bibfnamefont {G.~D.}\
  \bibnamefont {Lin}}, \bibinfo {author} {\bibfnamefont {L.-M.}\ \bibnamefont
  {Duan}}, \bibinfo {author} {\bibfnamefont {C.~C.}\ \bibnamefont
  {Joseph~Wang}}, \bibinfo {author} {\bibfnamefont {J.~K.}\ \bibnamefont
  {Freericks}},\ and\ \bibinfo {author} {\bibfnamefont {C.}~\bibnamefont
  {Monroe}},\ }\bibfield  {title} {\bibinfo {title} {Onset of a quantum phase
  transitions with a trapped ion quantum simulator},\ }\href@noop {} {\bibfield
   {journal} {\bibinfo  {journal} {Nat. {C}ommun.}\ }\textbf {\bibinfo {volume}
  {2}},\ \bibinfo {pages} {377} (\bibinfo {year} {2011})}\BibitemShut {NoStop}%
\bibitem [{Note5()}]{Note5}%
  \BibitemOpen
  \bibinfo {note} {In our estimation, numerical simulation using the full JCH
  model with more than 17 ion sites and the commensurate filling of one
  excitation per ion site requires memory that can hold a number of basis
  states as large as that in a 40-spin system and cannot be easily solved with
  existing classical computers.}\BibitemShut {Stop}%
\end{thebibliography}%


%

\end{document}